\documentclass{vldb}
\usepackage{booktabs} 
\usepackage{kbordermatrix}
\usepackage{algpseudocode} 
\usepackage{algorithm}	
\usepackage{hyperref}
\usepackage{xspace}
\usepackage{times}

\newtheorem{exmp}{Example}[section]
\newtheorem{remark}{Remark}

\newtheorem{definition}{Definition}
\newtheorem{lemma}{Lemma}
\newtheorem{theorem}{Theorem}
\newtheorem{corollary}{Corollary}

\newtheorem{fact}{Fact}[section]

\newcommand{\mrs}{\textsc{MargRS}\xspace}
\newcommand{\mht}{\textsc{MargHT}\xspace}
\newcommand{\mps}{\textsc{MargPS}\xspace}

\newcommand{\irs}{\textsc{InpRS}\xspace}
\newcommand{\iht}{\textsc{InpHT}\xspace}

\newcommand{\ips}{\textsc{InpPS}\xspace}
\newcommand{\iem}{\textsc{InpEM}\xspace}

\newcommand{\para}[1]{\smallskip \noindent {\bf #1}}
\newcommand{\parit}[1]{\smallskip \noindent {\em #1}}
\newcommand{\eat}[1]{}
\newcommand{\E}{\ensuremath{\mathsf{E}}}
\newcommand{\Var}{\ensuremath{\mathsf{Var}}}

\begin{document}
\title{Marginal Release Under Local Differential Privacy} 

\numberofauthors{2}
\author{
\alignauthor
  Graham Cormode, Tejas Kulkarni\\
\affaddr{University of Warwick}\\
\email{\{G.Cormode,T.Kulkarni.2\}@warwick.ac.uk}  
\alignauthor  
  Divesh Srivastava\\
\affaddr{AT\&T Labs--Research}
\email{divesh@research.att.com}
}

\maketitle

\allowdisplaybreaks

\begin{abstract}
Many analysis and machine learning tasks require the availability of
marginal statistics on multidimensional datasets while providing
strong privacy guarantees for the data subjects.
Applications for these statistics range from finding correlations
in the data to fitting sophisticated prediction models.
%
In this paper, we provide a set of algorithms for materializing marginal
statistics under the strong model of local differential privacy.
We prove the first tight theoretical bounds on the accuracy of marginals
compiled under each approach, perform empirical evaluation to
confirm these bounds, and evaluate them for tasks such as modeling and
correlation testing.
Our results show that releasing information based on (local) Fourier
transformations of the input is preferable to alternatives based
directly on (local) marginals.
\end{abstract}

\section{Introduction}
\label{sec:intro}

Modern data-driven applications must guarantee a high level of privacy
to their users if they are to gain widespread acceptance. 
The current de facto standard for privacy is differential privacy,
which imposes a statistical requirement on the output of a data
release process.
Considerable effort has been invested into achieving this guarantee
while maximizing the fidelity of the released information, typically
with the assistance of a centralized trusted third party who
aggregates the data. 
However, there is growing importance placed on 
algorithms which dispense with
the trusted aggregator, and instead allow each participant to ensure
that the information that they reveal already meets the differential
privacy guarantee in isolation.
This gives the {\em local differential privacy} (LDP) model.

The model of Local Differential Privacy combines the statistical
guarantees of differential privacy with a further promise to the user: their
information is never visible to anyone else in its raw form, and they
retain ``plausible deniability'' of any sensitive information
associated with them.
Local differential privacy has been adopted in Google Chrome 
via the RAPPOR tool to collect browing and system statistics~\cite{rappor:15},
and in Apple's iOS 10 to collect app usage statistics~\cite{appledp:17}.
It is consequently already deployed for information gathering over a
user base of hundreds of millions of people. 
So far, work in the LDP model has focused on relatively simple
computations: collecting basic frequency statistics about one
dimensional data (e.g. identifying popular destinations on the web).
But this one-dimensional view of the world does not capture the rich
set of correlations that real data exhibits.
The following case study gives some examples. 

\begin{figure}[t]
\centering
      \centering
\begin{tabular}{|c||c|c|c|c|}
\hline 
\rule[-1ex]{0pt}{2.5ex} User/Preferences & Action & Romance & Comedy & Crime \\ 
\hline 
\hline
\rule[-1ex]{0pt}{2.ex} Alice & Y & N & N & Y \\ 
\hline 
\rule[-1ex]{0pt}{2.5ex} Bob & N & Y & Y & N \\ 
\hline 
\rule[-1ex]{0pt}{2.5ex} $\vdots$ & $\vdots$ & $\vdots$ & $\vdots$ & $\vdots$ \\ 
\hline 
\rule[-1ex]{0pt}{2.5ex} Zayn & Y & N & Y & N \\ 
\hline 
\end{tabular}
\caption{User Preferences}
\label{fig:fulldist}
\end{figure}

\begin{figure}[t]
\centering
\begin{tabular}{|c|c|c|}
\hline 
Animation/Musical & Y & N \\ 
\hline 
Y & 0.55 & 0.15 \\ 
\hline 
N & 0.10 & 0.20 \\ 
\hline 
\end{tabular}
  \caption{$2$-way marginal}
  \label{fig:2way}
\end{figure}

\eat{
\begin{figure}
\centering
\begin{tabular}{|c|c|c||c|}
\hline 
Action & Romance & Crime & Fraction \\ 
\hline 
\hline
N & N & N & 0.04 \\ 
\hline 
N & N & Y & 0.12 \\ 
\hline 
\vdots & $\vdots$& $\vdots$ & $\vdots$ \\
\hline 
Y & Y & N & 0.17 \\  
\hline 
Y & Y & Y & 0.2 \\ 
\hline 
\end{tabular} 
\caption{A $3$-way marginal}
\label{fig:3way}
\end{figure}
}

\para{Motivating example: video viewing data.}
Consider collecting data on video views from a large user base.
Online video viewing habits should be considered very private, as
personal preferences can reveal highly sensitive information
about an individual's views on politics, religion and sexuality.
There is considerable precedent for considering such information in
need of protection, from the US Video privacy protection act, to the
debate around the data released as part of the Netflix
Prize~\cite{Narayanan:Shmatikov:06}.
Figure~\ref{fig:fulldist} shows an example data set in the context of
video preferences, where each user is described in terms of a number
of (binary) attributes, capturing which genres they have expressed
interest in. 

There are many potential uses of this kind of data when it is
represented as statistical marginal tables. 
Put simply, a marginal table records the (empirical) probability
distribution between a set of attributes.
A video service provider could learn which individual genres are more
or less popular within this user population.
They could also understand which combinations of genres are jointly
popular, to guide the procurement and promotion of new content --- for example,
\textit{generally, do viewers interested in watching animated movies also watch musicals?} 
We would also want to test the validity of any correlations found by
subjecting them to statistical hypothesis tests. 
Beyond just finding correlations, we may be interested in more complex
analysis such as describing the probabilistic relationship between
cause and effects by modeling features as a Bayesian Network.
Such graphical models rely on the computation of conditional
probability tables (CPTs), which are derived directly from marginals. 
Lastly, one could seek to design recommender systems based on the
correlations in the data: \textit{since you enjoyed these types of
  movie, we think you'll enjoy this one.}
All these tasks can be accomplished given information about the
population-level correlations between particular combinations of
attributes, which can be derived from the marginal distributions between small numbers of
attributes--typically, two-way or three-way marginals suffice. 
Figure~\ref{fig:2way} shows a marginal table between
musicals and animated films, which shows a
correlation between the categories. 
Consequently, our objective is to allow such tables to be computed
accurately on demand from data collected privately from a large user population. 
\qed

\smallskip
Thus the contingency, or marginal, table is the workhorse of data
analysis.
These statistics are important in and of themselves for understanding
the data distribution, and identifying which attributes are correlated.
They are also used for query planning and approximate query answering
within database systems.
Statistical inference and machine learning also rely on accurate
marginals capturing the correlations.
Consequently, almost any modern study of data requires the use of (low
order) marginals at some level.
So, we target low-order 2-way and 3-way marginals as
our main focus.

It is therefore not surprising that much work on private data analysis has studied
the problem of materializing and releasing marginals while achieving some privacy
guarantee.
It is clear that the information described by marginals is potentially
very sensitive, as it collates and reveals information about
individuals.
A canonical example of privacy leakage is when a cell in a marginal
table refers to just one person or a few individuals, and allows the value of an
attribute to be inferred.
For instance, a marginal table relating salary and zip code can reveal 
an individual's income level when they are the sole high earner in a region.
While this problem has been well-studied under centralized
differential privacy, there is limited prior work that applies in the local
DP model for two-way marginals~\cite{rappor2:16}; in this paper, we seek to give strong
guarantees for arbitrary marginals. 

\para{Our Contributions.}
In this work, we provide a general framework for marginal release
under LDP, with theoretical and empirical analysis.
First, we review prior work (Section~\ref{sec:prior}), and provide
background on methods to support private marginal release
(Section~\ref{sec:prelims}). 
We describe a set of new algorithms that give unbiased estimators for
marginals, which vary on fundamental design choices such as whether to
release information about each marginal in turn, or about the whole
joint distribution; and whether to release statistics directly about
the tables, or to give derived statistics based on (Fourier) transforms of the data.
For each combination, we argue that it meets the LDP guarantee, and
provide an accuracy guarantee in terms of the privacy parameter
$\epsilon$, population size $N$, and also the dimensionality of the data, $d$, and
target marginals, $k$ (Section~\ref{sec:algorithms}).
We perform experimental comparison to augment the theoretical
understanding in Section~\ref{sec:expts}, focusing mostly on the
low-degree marginals that are of most value.
Across a range of data dimensionalities and marginal sizes,
the most effective techniques are based on working in the Fourier (Hadamard)
transform space, which capture more information per user than methods
working directly in the data space.
The use of Hadamard transform for materializing marginals was
considered by early work in the centralized differential privacy
model, but has fallen from favor in the centralized model, supplanted
 by more involved privacy mechanisms~\cite{dwjz:11,hlm:12,tuv:12}.
We observe that these other mechanisms do not easily translate to the local
model.
However, methods involving the linear (Hadamard) transformation do so
more smoothly, yielding effective LDP protocols; 
our contribution here is in adapting, analyzing, and evaluating this
approach in the local model.
%
The endpoint of our evaluation is the application of our methods to
two use-cases: 
building a Bayesian model of the data, and testing statistical significance of correlations.
These confirm that in practice the Hadamard-based approach is
preferable and the most scalable. 
Our methods are eminently suitable for implementation in existing
LDP deployments  (Chrome, iOS) for gathering correlation statistics.

\section{Related Work}
\label{sec:related}
\label{sec:prior}
Differential Privacy (DP) \cite{dinur:03,dwork:04,dmns:06},
unlike its precursor privacy definitions, provides semantic mathematical promises on
individuals' privacy, interpreted as statistical properties of the
output distribution of a randomized algorithm.
Formally, an algorithm $\mathcal{A}$ meets the $\epsilon$-DP guarantee if, over pairs
of inputs $D, D'$ that differ in the presence of a single individual,
its output $\mathcal{A}(D)$ satisfies
\begin{equation}
\label{eq:dp}
 \frac{\Pr[ {\mathcal A}(D) =x ]}{\Pr[{\mathcal
 A}(D')=x]} \leq \exp(\epsilon),
 \end{equation}
where $x$ is any permitted output. 
The model has risen in popularity and adoption compared to earlier
attempts to codify privacy (such as $k$-anonymity and $\ell$-diversity
\cite{Samarati:Sweeney:98,Li:Li:Venkatasubramanian:07}), which can
leak sensitive information. 
It has been a topic of inquiry for diverse research communities
including
theory~\cite{v:17}, data management~\cite{mhh:16}, machine
learning~\cite{sc:13}, and systems/programming languages~\cite{n:15}.
We focus on two directions: the emergent model of {\em
  local} differential privacy (Section~\ref{sec:priorldp}), and private marginal release  (Section~\ref{sec:priormarginal}).

\subsection{Local Differential Privacy (LDP)}
\label{sec:priorldp}

Initial work on differential privacy assumed the participation of a \textit{trusted aggregator}, who curates the private
information of individuals, and releases information through a DP
algorithm. 
In practice, individuals may be reluctant to share private
information with the central data curator.
Local differential privacy instead captures the case when each user
independently (but collaboratively) releases information on their
input through an instance of a DP algorithm.
This model was first suggested by Evfimievski {\em et
  al.}~\cite{Evfimievski:Srikant:Agrawal:Gehrke:02} under the name of
$\gamma$-amplification, with an application to mining association
rules. 
Duchi {\em et
  al.}~\cite{d:13} studied a generalization of that model as a local
version of DP, and proposed a minimax framework with information
theoretic bounds on utility.

The canonical LDP algorithm is randomized response (RR), first developed in the
context of survey design in the 1960s~\cite{Warner:65}.
Users who possess a private bit of information flip it with some
probability $p$ to have plausible deniability of their response.
Collecting enough responses shared through this mechanism allows an
accurate estimate of the population behavior to be formed.
Randomized response is at the heart of many recent LDP algorithms,
most famously Google's deployment of
RAPPOR~\cite{rappor:15},
where RR is applied to a Bloom filter encoding a large set of possible
URLs. 
In a follow-up paper, Fanti {\em et al.}~\cite{rappor2:16} extend
RAPPOR's ability to identify strings which are frequent across the
user distribution, building them up piece by piece. 
However, their solution is somewhat specific to RAPPOR's case and does
not offer any guarantee on the accuracy. 

There is a growing theoretical understanding of LDP.
Kairouz {\em et al.}~\cite{ksv:14,ksv2:14} study how to 
estimate the frequency of a single categorical attribute,
and propose optimal generalizations of randomized response.
Closest to our interest is work on generating a histogram under
LDP, or identifying the peaks in the input (heavy hitters). 
This can be viewed as the problem of estimating a one-dimensional
marginal distribution. 
The state of art asymptotic lower bound and matching algorithm are due
to Bassily and Smith \cite{sb:15}.
They adapt ideas from dimensionality reduction (i.e. the
Johnson-Lindenstrauss lemma) to build a primitive to estimate the
weight of a single point in the distribution; this is used to find all
heavy hitters.
Qin {\em et al.} adapt this approach to the related problem of
identifying heavy hitters within set valued data~\cite{qykxk:16}. 
Most recently, Nguy{\^{e}}n {\em et al.} describe a general approach
for data analysis under LDP with multiple rounds~\cite{Xiao:16}.
They propose an orthogonal measurement basis which is
isomorphic to the Hadamard transform.
However, it is only used for one-dimensional data, rather than
the multidimensional data we consider. 

\subsection{Publishing Marginals Privately}
\label{sec:priormarginal}

Marginal tables arise in many places throughout data
processing.
For example, an OLAP datacube is the collection of all possible
marginals of a data set.
Consequently, there has been much work to release individual marginals
or collections of marginals under privacy guarantees.
To the best of our knowledge, these all assume the trusted aggregator
model.
We discuss a representative set of approaches, and whether they can be applied under LDP.

\para{Laplace Noise.}
The baseline for differential privacy is the sensitivity and noise
approach: we bound (over all possible inputs) the ``sensitivity'' of a
target query in terms of the amount by which the output can vary as a
function of the input.
Adding noise from an appropriate distribution (typically Laplace)
calibrated by the sensitivity guarantees privacy.
This approach transfers to LDP fairly smoothly, since the sensitivity
of a single marginal on $N$ users is easy to bound by $O(1/N)$~\cite{dmns:06}. 
A variant is to apply this release to a transformation of the data,
such as a wavelet or Fourier transform~\cite{Xiao:Wang:Gehrke:10,barak:07}.
Our contribution in this work is to refine and analyze the best way to
release marginals via transformations under the related guarantee of LDP. 

\para{Subset Marginal Selection.}
When the objective is to release many marginals --- say, the entire
data cube --- the above approach shows its limitations, since the
sensitivity, and hence the scale of the noise grows exponentially with
the number of dimensions: $2^d$. 
Ding et al.~\cite{dwjz:11} suggest computing low dimensional marginals
by selecting a subset of high dimensional marginals that add the least
amount of noise by posing subset selection as a constrained
optimization problem and propose a greedy approximation.
This solution does not translate naturally to the local model, since each
user has access to only her record and may come up with a different
subset locally compared to others.

\para{Multiplicative Weights.}
Several approaches use the \textit{multiplicative weight update
  method} to iteratively pick an output
distribution~\cite{hlm:12,hr:10,ghru:13}.
For concreteness, we describe a non-adaptive approach
due to Hardt {\em et al.}~\cite{hlm:12}.
The method initializes a candidate output uniform marginal, and
repeatedly modifies it so that it is a better fit for the data. 
To ensure DP, it uses the exponential  mechanism \cite{mt:07} to
sample a $k$-way marginal whose projection at a certain point in the
true data is far from the corresponding value for the candidate. 
The candidate is then scaled multiplicatively to reduce the
discrepancy. 
The sampling and rescaling step is repeated multiple times, and the
convergence properties are analyzed.
The number of steps must be limited, as the ``privacy budget'' must be
spread out over all steps to give an overall privacy guarantee. 
Applying the exponential mechanism in this way does not obviously extend to the LDP model.
In particular, every user's single input is almost equally far from any
candidate distribution, so it is hard to coordinate the sampling to
ensure that the process converges.
A natural implementation would have many rounds of communication,
whereas we focus on solutions where each user generates a single
output without further coordination. 

\para{Chebyshev polynomials.}
Thaler {\em et al.} consider  viewing a dataset as a
  linear function on marginals, and represent each record of a dataset via
  a $\gamma$-accurate Chebyshev polynomial~\cite{tuv:12}.
 Privacy is achieved by perturbing their coefficients.
The focus of the work is to speed up compared to multiplicative
weights solutions which need to range over an $O(2^d)$-sized representation
of the data.
This approach could plausibly be adapted to LDP, although the steps
required are far from immediate. 


\medskip
In summary, the LDP requirement to perturb elements of every single
record independently (which are sparse in our case) while preserving
the underlying correlations is not yet met by prior work, and so we
must give new algorithms and analyses.   

\section{Model And Preliminaries}
\label{sec:prelims}
In line with prior work~\cite{barak:07}, our main focus is on data
represented by binary variables.
This helps to keep the notation uniform, and highlights the key
challenges.
We discuss the modifications necessary to accommodate more general
attributes in Section~\ref{sec:nonbinary}.

In our setting, each user $i$ has a has a private bit vector $j_i \in
\{0,1\}^{d}$ that represents the values of the $d$ (sensitive)
attributes for $i$.
It is often more convenient to view the user's data instead as an
indicator vector $t_i$  of length
$2^{d}$ with $1$ at exactly one place $j_i$ and $0$'s at remaining
positions.
The domain of all such $t_i$'s is the set of identity basis vectors
$\mathcal{I}_{2^{d}\times 2^{d}}$.
This `unary' view of user data allows us to model the full contingency
table correspondingly as a vector (histogram) of length $2^{d}$ with each cell
indexed by $\eta\in \{0,1\}^{d}$ storing the count of all individuals
with that exact combination of attribute values. 

An untrusted aggregator (e.g.~a pollster) is interested in
gathering information on these attributes from the population of
users. 
Under the LDP model, the aggregator is not allowed (on legal/ethical
grounds) to collect any user $i$'s records in plain form.
The gathered data should allow 
running queries (e.g.~the fraction of users that use product $A, B$
but not $C$ together) over the interaction of at most $k \leq d$
attributes.
We do not assume that there is a fixed set of queries known a priori.
Rather, we allow arbitrary such queries to be posed over the
collected data. 
Our goal is to allow the accurate reconstruction of $k$-way
marginal tables under LDP.
We now formalize 
 Local Differential Privacy (Section~\ref{sec:ldp}), and
introduce examples and notation for computing marginals (Section~\ref{sec:marginals}).

\subsection{Local Differential Privacy}
\label{sec:ldp}


Local differential privacy (LDP) requires 
each data owner to perturb their
output to meet the DP guarantee.  
Any two tuples $t_i,t'_i \in \mathcal{I}_{2^{d}\times 2^{d}}$ are
considered {\em adjacent}, with $||t_i-t'_i||_1 = 2$ i.e. $t_i,t'_i$ are adjacent if they differ in their positions of $1$'s.
LDP upper bounds the ratio of probabilities of seeing the same outcome
for all adjacent tuples.
The definition is obtained by applying the differential privacy
definition of equation~\eqref{eq:dp} to a single user's data. 

\begin{definition}[$\epsilon$-local differential privacy (LDP)~\cite{d:13}]
A randomized mechanism $F$ is differentially private with parameter
$\epsilon$ iff for all pairs
of adjacent input tuples $t_i,t'_i \in \mathcal{I}_{2^{d}\times 2^{d}}$, and
every possible output tuple $R$ ($k \leq d$), we have
\begin{equation}
\label{eq:ldp}
\Pr[F(t_i)=R] \leq e^{\epsilon} \Pr[F(t'_i) = R]
\end{equation}   
\end{definition}

Because LDP requires each individual input to be perturbed, it is
expected to incur more noise overall than ``traditional'' DP models
where a trusted aggregator receives unperturbed inputs.
When we aggregate in the LDP model, the above definition ensures that
we cannot confidently distinguish whether $R$ is an outcome of
$F(t_i)$ or $F(t'_i)$, yielding plausible deniability to user $i$.
Note that under this definition
each user
reveals their presence in the input.
The model allows each user to operate with a different
privacy parameter, but for simplicity we state our results using a value
of $\epsilon$ which is shared by all users (in common with other work on LDP). 

%
%

\para{Basic Private Mechanisms.}
\label{subsec:building_blocks}
We describe primitives for LDP on simple inputs, 
which form building blocks for our protocols. 


\parit{{Randomized Response (RR):}}
We first formally define the classic mechanism for releasing a single
bit $b_i$ under privacy by having 
the user lie with some probability~\cite{Chaudhuri:88}.
In its simplest form, randomized response has 
each user $i$ report the true value of their input ($b_i$)
with probability $p_r > \frac12$.
Otherwise, $i$ gives the opposite response ($1-b_i$).
It is immediate that RR admits differential privacy with
$e^{\epsilon}=\frac{p_r}{1-p_r}$, by considering the probabilities of
the four combinations of input and output. 
Its simplicity has made it popular in practical systems~\cite{rappor:15,appledp:17,Xiao:16}. 


\parit{{Budget Splitting (BS) and Randomized Response with Sampling (RRS):}}
When each user holds $m$ pieces of information, a first approach
is to release information about all $m$ via a mechanism that
achieves $(\epsilon/m)$-LDP on each, thus effectively splitting the
``privacy budget'' $\epsilon$ (BS).
Standard composition results from the DP literature ensure that
 BS meets $\epsilon$-LDP~\cite{dmns:06}.
However, in general, accuracy is improved if
we instead sample one out of $m$ pieces of information and release this with
$\epsilon$-LDP~\cite{sb:15}, and this is confirmed by our
analysis of our protocols. 
In particular, if a user's information is represented as a binary vector
of dimension $m$, we can uniformly sample an index $j$ with
probability $p_s = 1/m$, and use Randomized
Response with parameter $p_r$ to release the value found there. 

\parit{{Preferential Sampling (PS) and Parallel Randomized Response
(PRR) for sparse vectors:}}
We often encounter cases where a user holds a sparse
vector: exactly one entry is 1, and the rest are 0.
The random sampling approach applied to the entries of the vector has
the disadvantage that most likely we will sample a zero entry, limiting the information revealed. 
We discuss two alternative approaches, which extend
randomized response in different ways.
The first
extension of RR
is a natural generalization also proposed e.g.~by \cite{ksv:14}, which
we call Preferential Sampling. 
Given a sparse vector $t \in \{0,1\}^m$ such that
$|t|=1$ and $t[j]=1$,
we sample an index $\ell$ according to the following distribution: 
\begin{equation*}
  \ell= 
\begin{cases}
        j: t[j]=1, & \text{with probability } p_s \\
        j':j'  \in_{R} [m] \setminus\{j\}, & \text{with probability } 1-p_s 
\end{cases}
\end{equation*}
In other words, we report the true index with probability $p_s$, while
each incorrect index is reported with probability $\frac{1-p_s}{m-1}$. 
When $m=2$ this mechanism is equivalent to  $1$ bit randomized response. 
Considering these two output probabilities, we immediately have: 
\begin{fact}
  The index $j$ reported by preferential sampling meets LDP with
  $e^{\epsilon} = \frac{p_s}{1-p_s}\cdot {m-1}$. 
\end{fact}
\noindent
Rearranging, we set $p_s = (1 + (m-1)e^{-\epsilon})^{-1}$ to
obtain $\epsilon$-LDP.

A second approach is apply $m$ independent instances of RR, each
with parameter $\epsilon/2$.
We denote this as Parallel Randomized response (PRR).
Note that the output of PRR is an $m$-bit string which is not
guaranteed to be sparse. 

\begin{fact}
Parallel Randomized Reponse applied to a sparse vector $t$ meets
$\epsilon$-LDP.
\end{fact}

To see this, observe first that the probability of seeing any
particular output string is the product of the
probabilities for $\frac{\epsilon}{2}$-RR applied to each bit of $t_i$
in turn.
Considering adjacent inputs $t_i$ and $t'_i$, and a particular output
$R$ in \eqref{eq:ldp}, the probabilities associated with all but two output bit locations in $R$
are identical (and so cancel in the ratio
$\Pr[F(t_i)=R]/\Pr[F(t'_i)]=R$).
We are left with the probabilities associated with the locations $j_i$
and $j'_i$ (i.e. where the 1 bits in the two adjacent inputs are).
The probability ratio for each of these bits is $\exp(\epsilon/2)$
from $\frac{\epsilon}{2}$-RR, so their product is
$\exp(\epsilon)$, as required by~\eqref{eq:ldp}.

\eat{
In other words, we inject some amount of precedence to the signal index $j$ with parameter $p'$ and sample $j$ with prob. $p'$. Otherwise, we sample a random index $j'$. It is worth noting that, in PS based methods, a user gains plausible deniability over index she reports and not over its value since we always report $1$. 
\begin{remark}
In the procedure described above, index $j$ has a second chance of getting sampled. One could also design variants of PS where $j$ does not get a second chance. Asymptotically these variants should perform similarly. 
\end{remark}
}

\subsection{Marginal Tables} 
\label{sec:marginals}
\noindent{\bf Notation and preliminaries.}
Recall that we model each user $i$'s bit vector $t_i \in \mathcal{I}_{2^{d}\times 2^{d}}$
as a vertex in a $d$-dimensional Hamming cube.
Then we can restrict our attention only on a subset of $k$ dimensions
of interest by summing (marginalizing) out cells of non-essential dimensions.
This is formally captured by the following definition. 
\begin{definition}[Marginal operator]
  Given a vector $t \in \mathbb{R}^{2^{d}}$, the {\em marginal operator}
  $\mathcal{C}^{\beta}: \mathbb{R}^{2^{d}} \Rightarrow \mathbb{R}^{2^{k}}$
  computes the summed frequencies for all combinations of
  values of attributes encoded by $\beta \in \{0,1\}^{d}$,
  where $|\beta|$, the number of $1$s in $\beta$, is $k \leq d$. 
\end{definition} 
%
For example,  for $d=4$ and $\beta=0101$
(which encodes our interest in the
second and the fourth attribute), the result of
$\mathcal{C}^{0101}(t)$ is the projection of $t$ on all possible
combinations of the second and fourth attributes with remaining
attributes marginalized out.
Each of the $2^k$ entries in the vector $\mathcal{C}^{0101}(t)$
stores the total frequency of combinations of the $k$
attributes identified by $\beta$. 
We make use of the $\preceq$ relation, defined as 
  $\alpha \preceq {\beta}$
  iff $\alpha \wedge \beta =\beta$.
%
%
For convenience of expression, we abuse notation and allow
$C^{\beta}(t)$
to be indexed by
$\{0,1\}^{d}$ rather than
$\{0,1\}^k$, with the convention that entries $\alpha$ such that
$\alpha \not\preceq \beta$
are 0. 
Under this indexing, the entries in a marginal can be written in the
following way: 
%
\begin{equation}
\textstyle
  \forall \gamma \preceq \beta \quad
C^{\beta}(t)[\gamma] = \sum_{\eta: \eta \wedge \beta = \gamma} t[\eta]
\label{eq:marginal}
\end{equation}
The condition $\eta \wedge \beta = \gamma$
selects all indices
$\eta \in \{0,1\}^{d}$ whose value on attributes encoded by $\beta$ are
$\gamma$.
%
\begin{exmp}
  \label{example}
Let $d=4$ and $\beta=0101$.
Then, applying~\eqref{eq:marginal}:\newline
\indent
$
C^{0101}(t)[0\mathbf{0}0\mathbf{0}] =
t[0\mathbf{0}0\mathbf{0}]+t[0\mathbf{0}1\mathbf{0}]+t[1\mathbf{0}0\mathbf{0}]+t[1\mathbf{0}1\mathbf{0}]
$\newline
\indent$
C^{0101}(t)[0\mathbf{0}0\mathbf{1}] =
t[0\mathbf{0}0\mathbf{1}]+t[0\mathbf{0}1\mathbf{1}]+t[1\mathbf{0}0\mathbf{1}]+t[1\mathbf{0}1\mathbf{1}]
$\newline
\indent$
C^{0101}(t)[0\mathbf{1}0\mathbf{0}] =
t[0\mathbf{1}0\mathbf{0}]+t[0\mathbf{1}1\mathbf{0}]+t[1\mathbf{1}0\mathbf{0}]+t[1\mathbf{1}1\mathbf{0}]
$\newline
\indent$
C^{0101}(t)[0\mathbf{1}0\mathbf{1}] =
t[0\mathbf{1}0\mathbf{1}]+t[0\mathbf{1}1\mathbf{1}]+t[1\mathbf{1}0\mathbf{1}]+t[1\mathbf{1}1\mathbf{1}]
$\newline
All indices in $\{0,1\}^{d}$ contribute exactly once to
one entry in $C^{0101}$. 
\end{exmp}

\begin{definition}[$k$-way marginals]
We say that $\beta$ identifies a $k$-way marginal when $|\beta|=k$.   
For a fixed $k$, the set of all $k$-way marginals correspond to all
${d \choose k}$ distinct ways of picking $k$ attributes from $d$.
We refer to the set of {\em full} $k$-way marginals as encompassing all $j$-way marginals sets, $\forall j \leq k$.
\end{definition}

Note that the (unique) $d$-way marginal corresponds to the complete input distribution. 
Since a single user's input $t_i$ is sparse i.e. contains just a
single $1$ (say at index $j_i$), any marginal $\beta$ of $t_i$
will also be sparse with just one non-zero element.
The relevant index in $\mathcal{C}^{\beta}(t_i)$
is given by the bitwise operation $j_i \wedge \beta$.  

\begin{definition}[Marginal release problem]
  \label{def:problem}
  Given a set of $N$ users, our aim is to collect information (with an
  LDP guarantee) to allow an approximation
  of any $k$-way marginal $\beta$ of the full $d$-way distribution
  $t = \sum_{i=1}^N t_i/N$. 
  Let $\widehat{\mathcal{C}^{\beta}}$ be the approximate answer.
  We measure the quality of this in terms of the {\em total variation
  distance} from the true answer $\mathcal{C}^\beta(t)$, i.e.
  \[
\textstyle
  \frac12 \sum_{\gamma \preceq \beta} | \widehat{\mathcal{C}^{\beta}}[\gamma]
  - \mathcal{C}^{\beta}(t)[\gamma] | = \frac12 \| \widehat{\mathcal{C}^{\beta}}
  - \mathcal{C}^{\beta}(t) \|_1\]
  \end{definition}


The marginals of contingency tables allow the study of
interesting correlations among attributes.
Analysts are often interested in marginals with relatively few
attributes (known as low-dimensional marginals).
If we are only concerned with interactions of up to at most $k$
attributes, then it suffices to consider the $k$-way marginals, rather
than the full contingency table. 
Since during the data collection phase we do not know a priori which
of the $k$-way marginals may be of interest, our aggregation should
gather enough information from each user to evaluate the set of full $k$-way
marginals for some specified $k$.
Our aim is to show that we can guarantee a small total variation
distance with at least constant probability\footnote{All our methods
allow the probability of larger error to be made arbitrarily small.}.
We will express our bounds on this error in terms of the relevant
parameters $N$, $d$, $k$, and the privacy parameter
$\epsilon$. 
To facilitate comparison, we give results using the $\tilde{O}$
notation which suppresses factors logarithmic in these parameters.

\para{Marginals and Basis Transforms.}
Since the inputs and marginals of individual users are sparse, the
information within them is concentrated in a few locations.
A useful tool to handle sparsity and ``spread out'' the information
contained in sparse vectors is to transform them to a different orthonormal
basis.
%
There are many well-known transformations which offer
different properties, e.g 
Taylor expansions, Fourier Transforms, 
Wavelets, Chebyshev polynomials, etc.
Among these, the discrete Fourier transformation over the Boolean
hypercube---known as the Hadamard transform---has many attractive
features for our setting.  
\begin{definition}[Hadamard Transformation (HT)]
\label{def:ht}
The transform of vector $t \in \mathbb{R}^{2^d}$ is $\theta = \phi t$
where $\phi$ is the orthogonal, symmetric $2^{d} \times
2^{d}$ matrix where
  $\phi_{i,j} = 2^{-d/2}(-1)^{\langle i \wedge j  \rangle}$.
\end{definition}
Consequently, each row/column in $\phi$ consists of entries of the
form $\pm \frac{1}{2^{d/2}}$, where the sign is determined by number
of bit positions that $i,j$ agree on.
It is straightforward to verify that any pair of rows $\phi_i, \phi_j$
satisfy $\langle \phi_i , \phi_j \rangle = 1$ iff $i=j$, and the inner
product is 0 otherwise.
Hence $\phi$ is an orthonormal basis for $\mathbb{R}^{2^d}$. 
Given an arbitrary vector $t$, we say that its representation under
the HT is given by the $2^{d}$ {\em Hadamard coefficients} (denoted as
$\theta$) in the
vector $\theta = \phi t$. 
These properties of HT are well-known due to its role in the theory of Boolean
functions~\cite{ABF:14}.
In our case when $t_i$ has only a single $1$ (say at index $\ell$), the
Hadamard transform of $t_i$ amounts to selecting the $\ell$th basis
vector of $\phi$, and so
$\theta_j = \phi_{j,\ell}$. 
We rely on two elements to apply the Hadamard transform in our setting.
The first follows from the fact that the transform is linear:
\begin{lemma}
\label{fact:sumcoeffs}
$
\phi (\sum_{i =1 }^{n} t_i/N) = { \frac1N} \sum_{i=1}^{n} (\phi t_i)$
\end{lemma}

That is, the Hadamard coefficients for the whole population are formed
as the sum of the coefficients from each individual.
The second ingredient  due to  Barak et al.~\cite{barak:07} is that we can write any marginal $\beta \in
\mathcal{C}$ as a sum of only a few Hadamard
coefficients.

\begin{lemma}[\cite{barak:07}]
\label{fact:ht}
Hadamard coefficients $H_k\!=\{\theta_\alpha\!:\!|\alpha|\!\le\!k\}$
are sufficient to evaluate any $k$-way marginal $\beta$.
Specifically,
\begin{equation}
  \mathcal{C}^\beta(t)_\gamma =
\sum_{\alpha \preceq \beta} \langle \phi_\alpha, t \rangle 
\sum_{\eta : \eta \wedge \beta =\gamma} \phi_{\alpha,\eta}
= 
\sum_{\alpha \preceq \beta} \theta_\alpha  \Big(\sum_{\eta : \eta \wedge \beta =\gamma}\phi_{\alpha,\eta}\Big)
\label{eq:htsum}
\end{equation}
\end{lemma}
%
%
Considering Example~\ref{example}, to compute the marginal
corresponding to $\beta = 0101$, we just need the four Hadamard
coefficients indexed as $\theta_{0000}, \theta_{0001}, \theta_{0100}$
and $\theta_{0101}$.
Moreover, to evaluate {\em any} 2-way marginal from $d=4$, we just
need access to the ${4 \choose 0} + {4 \choose 1} + {4 \choose 2} =
11$ coefficients whose indices have at most 2 non-zero bits, out of
the $2^4 = 16$ total coeffcients. 

\section{Private Marginal Release}
\label{sec:algorithms}

We identify a
number of different algorithmic design choices
for marginal release under LDP.
By considering all combinations of these choices, we reach a
collection of six distinct baseline algorithms, which we evaluate
analytically and empirically, and identify some clear overall preferred
approaches from our subsequent study. 
We describe our algorithms in terms of two dimensions:

\parit{View of the data.}
The first dimension is to ask what view the algorithm takes of the
data.
We are interested in marginals, so one approach is to project the data
out into the set of marginals of interest, and release statistics
about those marginals.
However, since any marginal can be obtained from the full input
distribution by aggregation, it is also possible to work with the data
in this form.  

\parit{How the information is released.}
The canonical way to release data under LDP is to apply Randomized
Response.
As discussed in Section~\ref{subsec:building_blocks}, when the user's
data is represented as a sparse input vector, we can instantiate this
by sampling a cell in their table, and applying Randomized
Response (the randomized response with sampling, RRS, approach);
by using a method such as preferential sampling approach (PS); or
parallel randomized response (PRR) to report information on the vector.
The alternative approach we study is to apply the Hadamard transform:
the user's table is now represented by a collection of coefficients,
each of which can take on one of two possible values.
We can then sample one Hadamard coefficient, and report it
via randomized response (we call this the HT approach).
Note that it is not meaningful to apply preferential sampling or
parallel RR after
Hadamard transform, since the input no longer meets the necessary sparsity
assumption. 

\smallskip
In order to analyze our algorithms, we make use of bounds from
statistical analysis, in particular (simplified forms of)
the Bernstein and Hoeffding inequalities:

\begin{definition}[Bernstein and Hoeffding inequalities]
  \label{def:bernstein}
  \label{def:hoeffding}
  Given $N$ independent variables $X_i$ such that
  $\E[X_i] = 0$, $|X_i| < M_i$, and $\Var[X_i] = \sigma^2$ for all $i$.
  Then for any $c>0$, 
  \[\textstyle
  \Pr\Big[\frac{|\sum_{i=1}^{N} X_{i} |}{N} > c\Big] \leq
\begin{cases}
  2\exp(-
  \frac{Nc^2}{2\sigma^2 + \frac{2c}{3}\max_i M_i})
& \text{(Bernstein ineq.)} \\
  2\exp(- \frac{N^2c^2}{2\sum_{i=1}^{N} M_i^2})
  &\text{(Hoeffding ineq.)}
\end{cases}
\]
\end{definition}

These two bounds are quite similar, but Bernstein makes
greater use of the knowledge of the variable distributions, and
leads to stronger bounds for us when we can show $\sigma^2 < M
= \max_i M_i$.

\subsection{Accuracy Guarantees}
\label{sec:accuracy}
In order to apply the above algorithms, we need to show that the
combination of techniques provides results which are unbiased,
accurate, and private.
We now prove that these properties hold for the baseline operators,
and go on to show what results hold for each method in turn. 
We first show how to obtain unbiased estimates for quantities of
interest from the privacy primitives. In each case, we apply the primitive to
estimate a parameter of the population $f$, such that $0 \le f \le 1$, using $N$ observations.  

\para{{Unbiased estimator for Preferential Sampling.}} 
Recall that preferential sampling reports an index $\ell$ which
is claimed to be the location of the sole 1 in the given input table
of size $2^k$. 
For convenience, we write $D=2^{k}-1$.
Given the $N$ reports, let $F_j$ be the fraction of times that $j$ is
reported, and let $f_j$ be the true fraction of inputs that have a 1 at location
$j$.
Then the expected reports of $j$ come from the
proportion of times the input is $j$ and it is correctly reported
($p_s f_j$), plus the proportion of times the input is not $j$ but $j$
is chosen to be reported ($(1-f_j) \frac{1-p_s}{D}$).  Thus
\centerline{$
\E[F_j] = p_s f_j + (1-p_s) \frac{1-f_j}{D}
$}
Rearranging this provides an unbiased estimator,
%
$
\hat{f}_j=
\frac{D F_j +  p_s -1 }{D p_s +  p_s -1}
$.
\eat{

\begin{lemma}
$\hat{f}_i$ is an unbiased estimator of $f_i$ i.e $\mathbb{E}[\hat{f}_i]=f_i$.

\end{lemma}
\begin{proof}
$\mathbb{E}[\hat{f}_i]=\mathbb{E}[\frac{DO_i+1-p'}{p'(D+1)-1}]=\frac{D\mathbb{E}[O_i]+1-p'}{p'(D+1)-1}=\frac{D\frac{\sum_{j=1}^{N}o^j_i}{N}+1-p'}{p'(D+1)-1}=\frac{D(p'f_i+\frac{(1-p')(1-f'_i)}{D})+1-p'}{p'(D+1)-1}=\frac{f'_i(Dp'-1+p')}{p'D+p'-1}=f'_i$
\end{proof}
}

\para{Unbiased estimator for Randomized Response methods.}
RRS picks one out of $m$ possible locations to report on and applies
randomized response.
Using the same notation as above, let $F_j$ be the fraction of the reports
that select index $j$ which report a particular output (say, $+1$).
Recall the true value is reported with probability $p_r$, and the
other answer with probability $1-p_r$.
With a similar calculation to the above (which can be viewed as a
special case of PS when $D=1$), 
we obtain an unbiased estimator
$
  \hat{f_j} = \frac{F_j + p_r - 1}{2p_r - 1}
$.
Following the corresponding calculation, 
the same estimator also results for each bit position in PRR. 

\para{\textbf{Unbiased estimator for Hadamard Transform}.}
The HT approach applies the Hadamard Transform to the user's input.
To achieve LDP, we can use randomized response with parameter $p_h$
on a chosen coefficient. 
Now let $f_j$ be the true unknown fraction of negative
values of coefficient $j$ over the population.
The expected observed fraction of instances reported as negative,
denoted $F_j$, is
\begin{align*}
  \begin{array}{rl}
\E[F_j]\!\!\!\!&
= -p_h f_j-(1-f_j)(1-p_h)+ (1-p_h)f_j+(1-f_j)p_h
\\&  = -4pf_j-1+2p_r+2f_j.
\end{array}
\end{align*}

Rearranging gives an unbiased estimator of the population fraction of
negative coefficients as
$\hat{f}_j=\frac{1- 2p_h + F_j}{2(1-2p_h)}.$
This gives an unbiased estimate of the $j$th Hadamard coefficient
$\hat{\theta}_j$ as
%
\[\begin{array}{rl}
\hat{\theta}_j & = 2^{-\frac{d}{2}} ( 1 - 2\hat{f}_j)
  = 2^{-\frac{d}{2}}\Big(1-2\Big(\frac{1-2p_h+F_j}{2(1-2p_h)}\Big)\Big) \\
  & = 2^{-\frac{d}{2}}\frac{1-2p_h-F_j-1+2p_h}{1-2p_h}=
  2^{-\frac{d}{2}}\frac{ F_i}{2p_h-1}.
\end{array}\]



%

\para{Master Theorem for Accuracy.}
To analyze the quality of the different algorithms,
we provide a generalized analysis that can be applied to several of our
algorithms in turn.  
%
%
We assume that each user input is in  $\{-1,1\}$ in the proof,
but we will also be able to apply the theorem when
inputs range over other values.
%
\begin{theorem}
\label{thm:master}
Let each $t_i$ be a sparse vector
where one entry is $\{-1,1\}$, and the rest are zero.
When each user $i$ uniformly samples an input element
$j$ with probability $p_s$ and applies randomized response with $p_r$
to construct $t^*_i$,
for $c > 0$ we have 
\[
\Pr\left[\frac{|\sum_{i=1}^N t_i^{*}[j] -t_i[j]|}{N} \geq c\right] \leq 2\exp \Big(-\frac{Nc^{2}p_s(2p_r-1)}{2p_r(2\frac{1-p_r}{2p_r-1}+\frac{c}{3})}\Big)\]
\end{theorem}

Intuitively, this theorem lets us easily express the (total variation)
error in a marginal as a function of parameters $p_s$ and $p_r$.
We will choose values of $c$ that make this probability constant ---
this implies (for example) that $c$ should be chosen proportional to $1/\sqrt{N
  p_s}$.
Hence, we capture how the error decreases as $N$ increases,
and how it increases as the number of items being sampled from increases. 
For ease of reading, we defer all lengthy proofs to the Appendix. 

\subsection{Input Perturbation Based Methods}

The three approaches which work directly on the input data
require a two-step analysis: first we consider the accuracy of
reconstruction of some global information (e.g. the full
distribution), then we analyze the accuracy of aggregating this to
give the required marginal $\beta$.
Throughout we assume that
$2^d$ is at least $\tilde{O}(N)$,
i.e. the number of users $N$
participating is at least proportional to the number of cells in the
full distribution ($2^d$).
This is natural, since it requires our methods which sample cells from
the full input to have at least constant probability of probing any
given cell.
Now we spell out the details of our input perturbation based algorithms.

\eat{
For input perturbation based methods, entire marginal evaluation happens at the aggregator's side and users only transmit noisy versions of their inputs to the aggregator. Upon collection, analyst has a choice to evaluate only a marginal of interest out of $\mathcal{C}$ using algorithm~\ref{alg:construct_marg_barak}. }

\para{Parallel Randomized Response On Input (\irs).}
Each user $i$ perturbs their value $t_i$ at every index
$\ell \in 2^d$ using $\frac{\epsilon}{2}$-RR (PRR) to obtain $b^*_{i,\ell}$ 
and releases the tuple $\langle \ell,b^{*}_{i,\ell} \rangle$.
We reconstruct a version of the full input $t^*$ by simply summing all
these contributions (and dividing by $N$); any desired marginal
$\beta$ is obtained by taking
$\mathcal{C}^\beta(t^*)$, i.e. computing that marginal of the
reconstructed input. 
This is the most direct application of LDP to this problem,
but may not yield the best accuracy, due to the high amount of noise
added in $2^d$ locations.
It is also potentially costly to apply, since each user needs to
materialize and communicate $2^d$ pieces of information. 
Applying our general analysis allows us to
bound  the error (total variation distance) in
 the returned marginal.

\begin{theorem}
\label{thm:irs}
\irs achieves $\epsilon$-LDP and guarantees that \newline
$\| \mathcal{C}^\beta(t)- {\mathcal{C}^{\beta}(t^*)}\|_1 =
\tilde{O}\Big(\frac{2^{(d+k)/2}}{\epsilon\sqrt{N}}\Big)$
with constant probability. 
\end{theorem}

\noindent{\bf Preferential Sampling On Input (\ips).}
As the name suggests, instead of applying binary randomized response, each $i$ in \ips
samples the input signal index $j$ with probability $p_s$, then reports the
selected index to the aggregator.
The reconstructed distribution $t^*$ is found by averaging all these
(unbiased) reports.
As in the previous case, we can obtain any desired marginal by
aggregating the reconstructed distribution. 
Then we prove the following theorem. 


\begin{theorem}
  \label{thm:psmaster}
\ips achieves $\epsilon$-LDP and guarantees that
  with constant probability
we have for a target $k$-way marginal $\beta$
\[
\textstyle
\| \mathcal{C}^\beta(t) - \mathcal{C}^\beta(t^*) \|_1 = \tilde{O}\left({\frac{2^{d+k/2}}{\epsilon\sqrt{N}}}\right).
\]

\end{theorem}
Consequently, we get a guarantee for \ips
in terms of total variation distance
of $\tilde{O}\left(\frac{2^{k/2}2^d}{\epsilon\sqrt{N}}\right)$.
This exceeds the bound of the previous algorithm by a factor of
$2^{d/2}$, so we expect the former to be more accurate in practice. 


\para{Random Sampling Over Hadamard Coefficients (\iht).}
In this method, user $i$ takes a randomly sampled coefficient $j$ from the
HT of her input (i.e. $\phi_i[j]$),  and releases it via Randomized
Response. 
According to Lemma~\ref{fact:ht}, we do not need to sample from all coefficients;
rather, we need only those coefficients required to reconstruct the
$k$-way marginals.
These are those coefficients whose $d$-bit (binary) indices contain at most
$k$ 1's.
There are $T = \sum_{\ell=1}^{k} {d \choose \ell} = O(d^k)$ of these,
which can be much smaller than all $2^d$ coefficients, or the $2^d$
parameters needed to describe the full input. 
The aggregator then builds an estimate for each coefficient as the
average of the reported values, and uses these to construct any target
marginal $\beta$ via the construction of Lemma~\ref{fact:ht} to
generate $\mathcal{C}^\beta(t^*)$.  
We can apply Theorem~\ref{thm:master} to this setting to obtain a
guarantee on the total variation distance between the correct and
reconstructed marginals. 
%
\begin{theorem}
\label{thm:iht}
  \iht achieves $\epsilon$-LDP, and with constant probability
we have for any target $k$-way marginal $\beta$
\[
\textstyle
\|\mathcal{C}^\beta(t) - \mathcal{C}^\beta(t^*)\|_1 = 
\tilde{O}\left(\frac{(2d)^{k/2}}{\epsilon \sqrt{N}}\right).\]
%
%
\end{theorem}

Comparing this to the previous results, we observe that the dependence
on ${2^{k/2}}\big/{\epsilon\sqrt{N}}$ is the same.
However, our full analysis shows a dependence on $\sqrt{T}$ in place of
$\sqrt{2^d}$.
Recall that $T = \sum_{\ell=1}^{k} {d \choose k} < 2^d$ for $k<d$.
For small values of $k$, this is much improved.
For example, for $k = 2$, $T=O(d)$ in \iht compared to a $2^{d/2}$ term for
\irs.
For completeness, in the appendix we present
Algorithms~\ref{alg:iht_user} and \ref{alg:input_aggr} to spell out
the steps followed by user and  aggregator in \iht.

\subsection{Marginal Perturbation Based Methods}

Our next methods are the analogs of the Input perturbation
methods, applied to a randomly sampled marginal rather than the full
input. 
For brevity, we omit the formal proofs of these results and instead
provide the necessary intuition, since they
are mostly adaptations of the previous proofs. 

\para{RR On A Random Marginal (\mrs).} 
In \mrs, user $i$ samples a random marginal $\beta_i \in \mathcal{C}$,
and then materializes all indices $j_i$ from within that marginal,
and perturbs each $b_i = \mathcal{C}^\beta(t)[j_i]$ using
$\frac{\epsilon}{2}$-RR (PRR).
Let $b^{*}_i$ denotes the bit obtained after perturbation at a
particular location.
The user then transmits the tuples $\langle \beta_i,j_i,b^{*}_i
\rangle$ to the aggregator, who simply sums the returned values and
normalizes for each marginal. 

\parit{Analysis.}
As with \irs, it is immediate that the method achieves $\epsilon$-LDP,
since only $b^{*}_i$ is specific to the input, and is obtained via RR
which is $\epsilon$-LDP.
In terms of accuracy, the analysis is also very similar to \irs.
The difference is that we are now considering
sampling from ${d \choose k}$ marginals, each of which contains $2^k$
pieces of information.
So where before we had a dependence on $2^d$, the method now also depends on
$1/p_s = {d \choose k}  = O( d^{k})$.
Thus, via Theorem~\ref{thm:irs}, we obtain a bound on the error in each entry of each marginal of
$\tilde{O}({\sqrt{d^k}}\big/{\epsilon\sqrt{N}})$.
Summing this over the $2^k$ entries in the marginal,
we obtain a total error of $\tilde{O}\left(\frac{2^{k} d^{k/2}}{\epsilon \sqrt{N}}\right)$. 

\para{PS On A Random Marginal (\mps).}
As an alternative approach to \mrs, we can use preferential sampling (Section~\ref{subsec:building_blocks})
to try to pick the entry in the randomly sampled marginal which
contains the 1, instead of materializing each entry independently
using randomized response.
For small marginals (i.e. small $k$), this may be effective. 
Otherwise the algorithm is similar to \mrs, and we build all the
required marginals by averaging together the (unbiased) reported
results from all participants. 


\parit{Analysis.}
The behavior of this algorithm can be understood by adapting the
analysis of
\ips.
Since we work directly with the marginal of size $k$, we now obtain a
bound in terms of $2^{3k/2}$ where before we had $2^{d+k/2}$. 
However, the effective population size is split uniformly across the
${d \choose k}$ different marginals.
Consequently, the total variation distance is
$\tilde{O}\left(\frac{2^{3k/2} d^{k/2}}{\epsilon\sqrt{N}}\right)$.
This exceeds the previous result by a factor of $2^{k/2}$, but for
small $k$ (such as $k=2$ or $k=3$), this can be treated as a constant
and the other factors hidden in the big-Oh notation may determine the true behavior. 


\para{Hadamard Transform Of A Random Marginal (\mht).}
\mht also deviates from \mrs only in how the chosen marginal is
materialized: it takes the Hadamard transform of each user's sampled marginal, and
uses RR to release information about a randomly chosen coefficient.
These are aggregated to obtain estimates of the (full) transform for
each $k$-way marginal $\beta$. 
Note that this method does not share information between marginals,
and so does not obtain as strong as result as \iht. 

\parit{Analysis.}
Here, $p_r$ is the same as in \iht, but we are now sampling over a
larger set of possible coefficients: each marginal requires $2^k$
coefficients to materialize, and we sample across $T = O(d^k)$
marginals.
This sets $p_s = O((2d)^{-k})$.
We obtain that
$\sigma^2 = O((2d)^k/\epsilon^2)$ and
$M = O((2d)^k/\epsilon)$.
Thus, we bound the absolute error in each reconstructed 
coefficient by $\tilde{O}\left(\frac{d^{k/2}}{\epsilon\sqrt{N}}\right)$,
by invoking Theorem~\ref{thm:master} with these values and then
applying the rescaling by $2^{-k}$.
We directly combine the $2^k$ coefficients needed by marginal
$\beta$, giving total error
$\tilde{O}(\frac{2^{3k/2}d^{k/2}}{\epsilon\sqrt{N}})$, similar to the
previous case. 

\para{Summary of marginal release methods.}
Although different in form, all three marginal based methods achieve
the same asymptotic error, which we state formally as follows: 
\begin{lemma}
  \label{lem:mrs}
Two marginal-based methods (\mps and \mht) achieve
$\epsilon$-LDP and with constant probability 
the total variation distance between true and reconstructed $k$-way marginals is at most
$\tilde{O}(\frac{2^{3k/2} d^{k/2}}{\epsilon \sqrt{N}})$.
For \mrs, the bound is 
$\tilde{O}(\frac{2^{k} d^{k/2}}{\epsilon \sqrt{N}})$.
\end{lemma}
Comparing to the input based methods, a dependence on a factor of
$\frac{2^{k/2}}{\epsilon\sqrt{N}}$ is common to all.
Marginal-based methods multiply this by a factor of at least $(2d)^{k/2}$,
while input based methods which directly materialize the full marginal
 (\irs and \ips) have a factor of $2^d$.
The input Hadamard approach \iht reduces this to just $d^{k/2}$.
Asymptotically, we expect \iht to have the best performance.
However, for the parameter regimes we are interested in (e.g. $k=2$),
all these bounds could be close in practice.
Hence, we evaluate the methods empirically to augment
these bounds. 

\subsection{Expectation-Maximization Heuristic}
\label{sec:iem} 
While materialization of marginals has not been the primary focus of
prior work, a recent paper due to Fanti {\em et al.}~does suggest an
alternative approach for the 2-way marginal case~\cite{rappor2:16}.
The central idea is for each user to materialize information on all
$d$ attributes, and to use post-processing on the observed
combinations of reported values to reach an estimate for a given
marginal. 

In more detail, each user independently perturbs each of the $d$
(binary) attributes via $(\epsilon/d)$-randomized response, i.e. using
Budget Sharing (BS).
To reconstruct a target marginal distribution, the aggregator applies
an instance of Expectation Maximization (EM). 
Starting from an initial guess (typically, the uniform marginal), the
aggregator updates the guess in a sequence of iterations.
Each iteration first computes the posterior distribution given the
current guess, applying knowledge of the randomized response mechanism
(expectation step).
It then marginalizes this posterior using the observed values of
combinations of values reported by each user, to obtain an updated
guess (maximization step).
These steps are repeated until the guess converges, which is then
output as the estimated distribution.
As noted by Bassily and Smith~\cite{sb:15}, this is a
plausible heuristic, but does not provide any worst case guarantees of accuracy.
We compare this method, denoted \iem, to the algorithms above in our
experimental study. 
In summary, we find that the method provides lower accuracy than our
new methods.
In particular, we see many examples where it fails: the EM procedure
immediately terminates after a single step and outputs the prior
(uniform) distribution.\footnote{Figure~\ref{fig:EM_fail} in the
  Appendix quantifies this in more detail, and shows some parameter
  settings where the method fails universally.}. 
We compare \iem with best of our methods in section~\ref{sec:Fanti_et_al}.

\newlength{\figsize}
\setlength{\figsize}{0.88\textwidth}

\begin{table}[t]
  \caption{Attributes of NYC taxi dataset}
  \centering
\begin{tabular}{|c|c|}
\hline
Attribute & Explanation \\
\hline 
\hline 
CC  &  Has customer paid using credit card? \\ 
\hline 
Toll & Has customer paid toll? \\ 
\hline 
Far & Is journey distance $\geq$ 10 miles? \\ 
\hline 
Night\_pick &  Is pickup time $\geq$ 8 PM? \\ 
\hline 
Night\_drop & Is drop off time $\leq$3 AM? \\ 
\hline 
M\_pick & Is trip origin within Manhattan? \\ 
\hline 
M\_drop &  Is trip destination within Manhattan? \\ 
\hline 
Tip &  Is tip paid $\geq$ 25\% of the total fare?  \\ 
\hline 
\end{tabular}
\label{tab:nyc_taxi} 
\end{table}

\begin{figure}[t]
\centering
  \includegraphics[width=0.9\columnwidth]{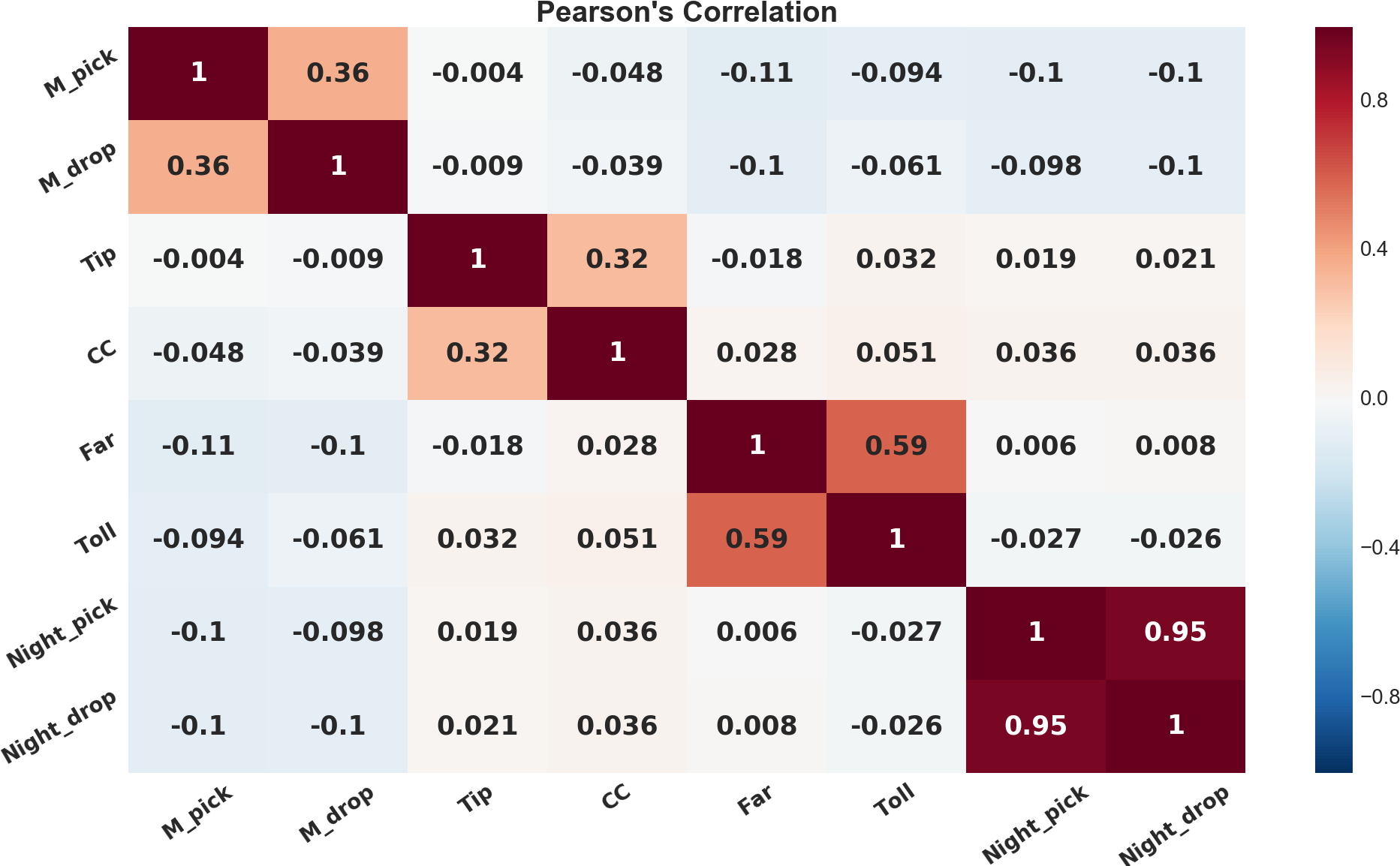}
  \caption{Heatmap of attribute correlation in NYC taxi dataset}
  \label{fig:corr}
\end{figure}

\begin{figure*}[t]
\centering
  \includegraphics[width=\figsize]{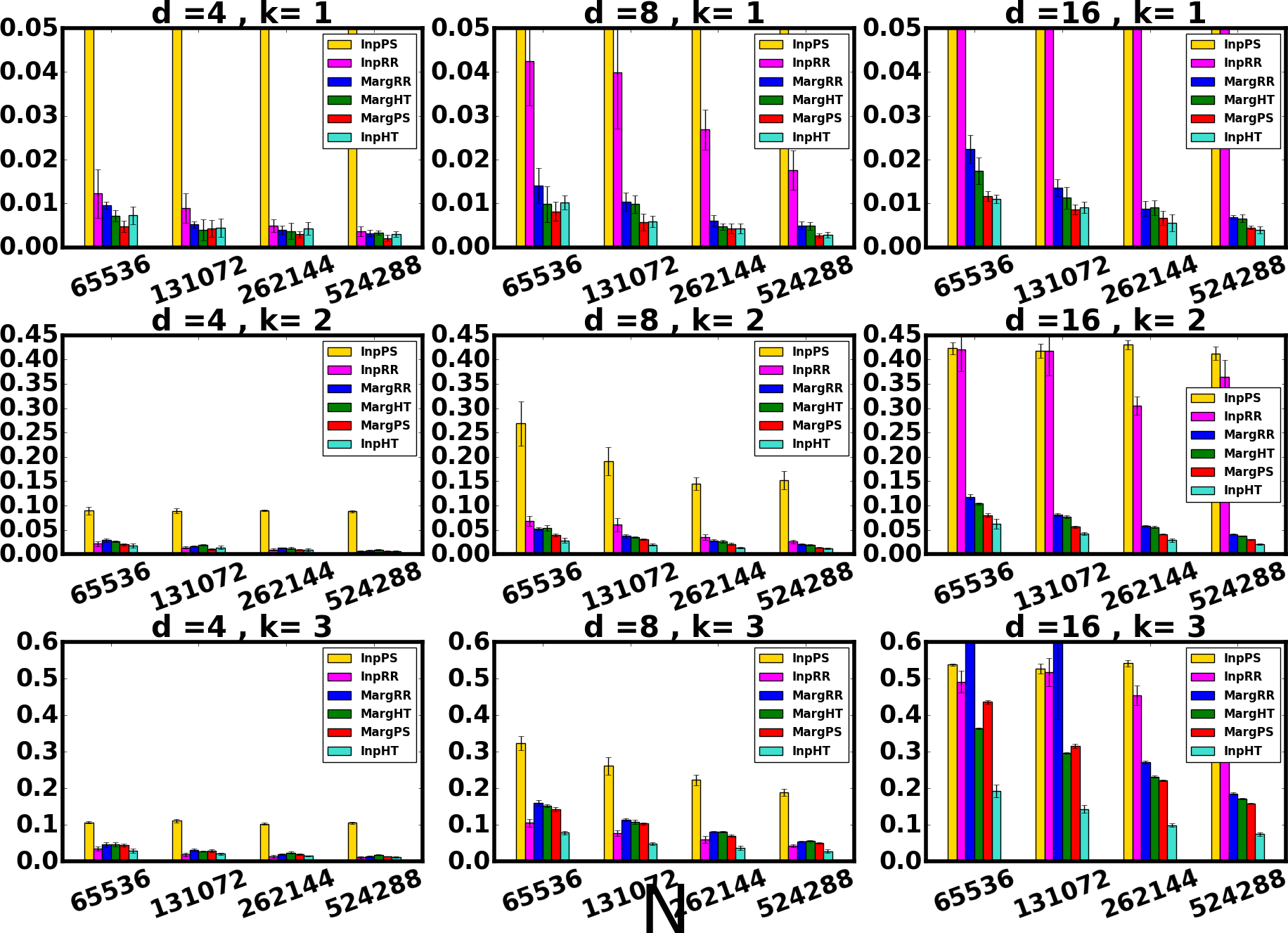}
  \caption{Mean total variation distance for $1,2,3-$way marginals
  over the movielens dataset as $N$ varies}
  \label{fig:l1_recon}
\end{figure*} 

\eat{
\section{On Marginals For large Parameter Regimes} It is apparent that we should be able to construct marginals using our methods under privacy for any value of $k$. However, it is not clear if  explicitly materializing a $k$-way marginal using our framework is the most preferred method for $k>2$ regimes. This is because our methods are data agnostic and weigh all attributes equally while adding noise. It may be possible to design better data dependent algorithms after understanding associations among variables in a dataset. We describe below a plausible approach.
\par One can approximate  a high dimensional joint distribution  as a product of multiple distributions containing at most two variables (cf.\ref{sec:chow_liu}). Once we learn the variables involved in each conditional probability term  required to best approximate a cell in a higher order marginal, we can  compute these terms using just $1,2$-way marginals and finally approximate a marginal without explicitly evaluating a it. Hence, we limit our experimental attention in the remainder of the paper to $1,2-$way marginals though we show some results for the $k=3$ case.    
}

\section{Experimental Evaluation}
\label{sec:expts}
We have two goals for our empirical study: (1) to give experimental
confirmation of the accuracy bounds proved above;
and
(2) to show that our algorithms support interesting machine
learning/statistical tasks using our marginal computing machinery as primitives.
We implement our experiments using standard Python packages (Numpy,
Pandas) and perform tests on a standard Linux laptop.

\begin{figure*}[t]
\centering
    \includegraphics[width=\figsize]{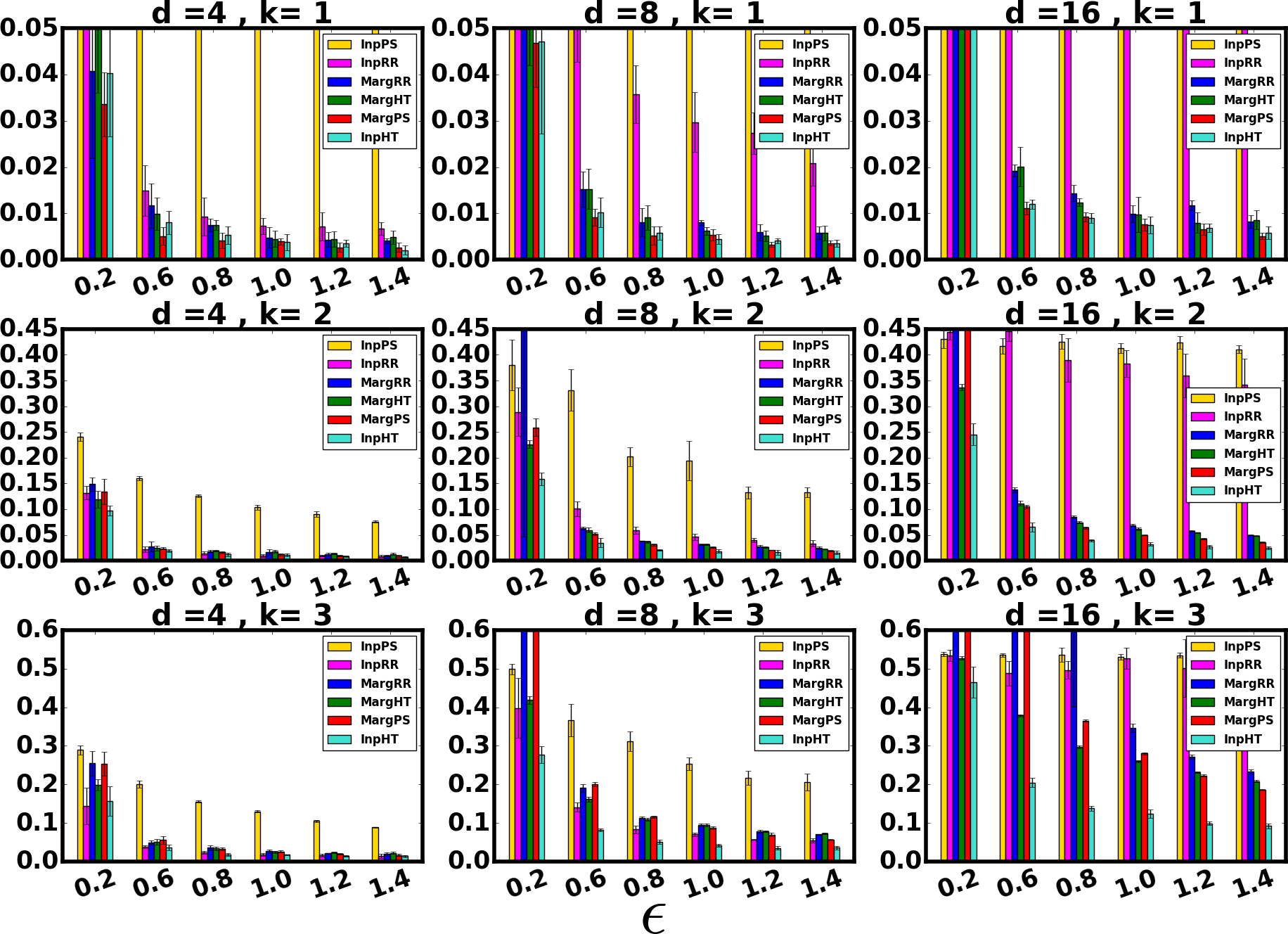}
  \caption{Mean total variation for $1,2,3-$way marginals for 
  $N=256K$ movielens users as $\epsilon$ varies. }
  \label{fig:vary_eps}
\end{figure*}

\subsection{Experimental Setting}

\para{Datasets used.}
We use two different datasets for our experiments:

\parit{Movielens \cite{hk:16}.}
This dataset comprises over $20M$ records from over $150K$ anonymous
users who rate nearly $40K$ unique movie titles. Each title belongs to
one or more of $17$ genres such as Action, Comedy, Crime, Musical
etc.
From this, we derive a dataset that resembles the ``video viewing''
example given in the Introduction. 
We first find the top-1000 most rated movies in each genre.
We assign each user a vector of preferences $t_i \in \{0,1\}^{d}$. For
each user $i$, a bit at index $j \in [d]$ is $1$ if $i$ has rated at
least one of the top $1000$ movies of genre $j$ and zero otherwise.
In this data, most attribute pairs are postively correlated. 

\parit{{NYC Taxi Data \cite{taxi:17}.}}
This dataset samples trip records from all trips completed in yellow
taxis in NYC from 2013-16.
Each trip record can be viewed a unique anonymous cab driver's
response to a set of survey questions about her journey.
Some of the attributes are GPS co-ordinates/timestamps of
pick-up/drop-off, payment method, trip distance, tip paid, toll paid,
total fare etc.
From this (very large) data set, we select out the $3M$ records having 
pickup and/or drop-off locations inside Manhattan.
We obtain 8 binary attributes for each trip, 
as detailed in Table~\ref{tab:nyc_taxi}. 
We observe in this dataset that most journeys are short, and so
attribute pairs such as pickup/drop-off locations/times, tip-fraction
and payment mode are strongly correlated.
Meanwhile, most other attribute pairs are negatively correlated, or
only weakly related. 
Figure~\ref{fig:corr} gives a heatmap for the
strength of pairwise associations using the Pearson coefficient.

\para{Default Parameters And Settings.} 
 In each experimental instance, we sample (with replacement) a set of
 random unique records/users ($50K \leq N \leq 0.5M$) as a power of
 $2$ from the total available population.
 We vary $\epsilon$ from $0.2$ (higher privacy) to $1.4$ (lower
 privacy).
Note that the theory shows that $\epsilon$ and $N$ are tightly
 related: decreasing $\epsilon$ means $N$ must be increased to obtain
 the same accuracy.
 Some prior work on LDP e.g. \cite{Xiao:16} studies
a smaller regime of $\epsilon$ values, at the expense of a much
 larger user population.
We begin our experimental study by sampling (without replacement) a small subset of
 dimensions $d$ (3-8), and increase to larger dimensionalities for our later experiments. 
Per our motivation (Section~\ref{sec:intro}), we focus on small
 marginals ($k=1,2,3$).  
We repeat each marginal reconstruction $10$ times to observe the
consistency in our results, and show error bars.


\begin{figure*}[t]
\centering
  \includegraphics[width=\figsize]{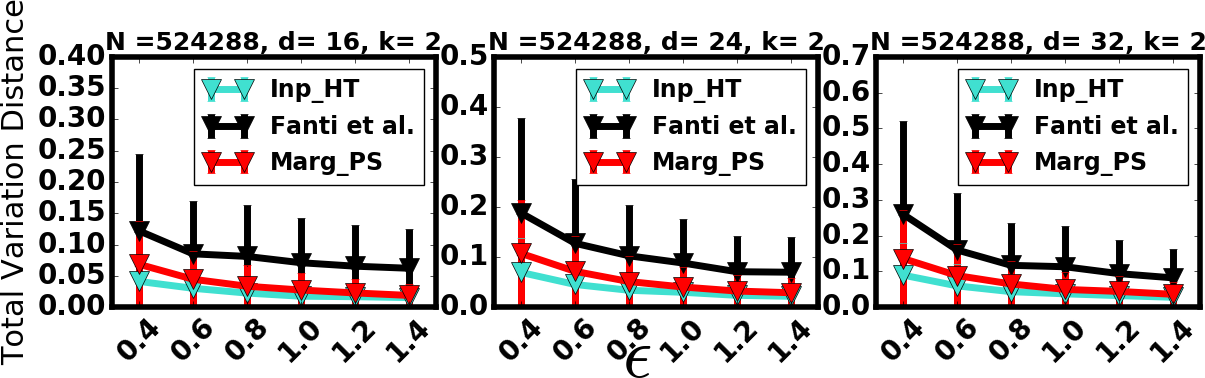}
 \caption{Total variation distance for $k=2$ on NYC Taxi Trips Data For larger $d$'s.
 We observe that our best algorithms outperform  Fanti et al.'s (\iem) algorithm for small $\epsilon$ values.}
  \label{fig:inp_ht_vs_inp_em_k2}
\end{figure*}

\subsection{Impact of varying population size $N$} 

We aim to understand how much a privately reconstructed marginal
$\mathcal{C}^\beta(t^{*})$
deviates from its non-private counterpart $\mathcal{C}^\beta(t)$
when $\beta$ is drawn from the set of $k$-way marginals.
First, we fix $\epsilon=1.1$ and vary $N$ for different choices of $d,k$.
For our initial comparison, we keep $d$'s moderate ($\{4,8,16\}$), as this
suffices to distinguish the methods which scale well from those that
do not.


\para{{Experimental Setting.}}
Figure~\ref{fig:l1_recon} shows plots for total variation distance in
reconstruction of $k$-way marginals
as we vary $N$ for all combinations of $k \in \{1,2,3\}$ and $d \in
\{4,8,16\}$
on the movielens
dataset with $\epsilon=\ln(3)\approx 1.1$ fixed throughout the
experiment.
Each grid point shows the mean variational distance of all
$k=1,2,3$ marginals.
The values of parameters $d$ and $k$ vary across the rows and columns
of the figure, respectively. 

\para{{Experimental Observations.}}
The first high level behavior to observe across the board is reduction
in error with increase of $N$ for all $6$ algorithms.
This agrees with the analysis that error should be proportional to
$1/\sqrt{N}$, i.e.~error halves as population quadruples. 
We also see an increase in error along columns (rows) as $k$ ($d$)
increases, although the dependency varies for different algorithms.  

Our second observation is that the performance
of \ips decays rapidly as a function of $d$, consistent with
the accuracy bound of $2^d$.
Typically, \ips's error does not reduce as with $N$.
This is because the probability of outputting the signal index becomes
so small for larger $d$'s that each user responds with a random index
most of the time. This means that the perturbed input distribution does not contain much information for our estimators to invert the added noise with precision.
One surrogate for the accuracy of the algorithms is the number of statistics materialized in each case. For $d=8, k=2$, \ips construct $2^8 = 256$ values, while the marginal-based methods are working on ${8\choose 2} \times 2^k = 112$ values.
As a result, the number of data points per cell is proportionately more for \mht, \mps thus improving their accuracy. 
On the other hand, the input-based method \iht convincingly achieves the lowest (or near lowest) error across all parameter settings. 


Breaking the algorithms down by the cardinality of the marginal ($k$),
note that for $k=1$ then the primitives RR and PS are effectively the
same.
Further, for a given marginal, there is only one meaningful Hadamard
coefficient needed, and so we expect the Hadamard-based methods to behave
similarly. 
Indeed, the methods \mps, \mrs, \mht, and \iht are largely
indistinguishable in their accuracy.
For the larger 2-way and 3-way marginals, we see more variation in behavior. 
The input-based methods do not fare well: \ips has very large errors for even
smaller $d$ values ($d=4$ and $d=8$), and \irs is similar once $d=16$.
We observe that \mps achieves better accuracy than \mrs.
This supports the idea that the former method, which preferentially
reports the location of each user's input value, can do better than
naive randomized response, even though this is not apparent from the
asymptotic bounds. 
Interestingly, on this data we see that the difference in
performance of \mps and \mht is tiny, and \mps turns out to be a better algorithm. For $d=16$, \mht starts as a better algorithm but is outperformed by \mps. 
is small compared to $d$ with increase in $N$.

\irs is among the better methods for smaller values of  $d$ and $
k$'s.
However, we advise against \irs for large $d$'s since it takes time
proportional to $2^d$ to perturb all cells of each user.
Similarly, the use of \mrs is also hard to justify from an execution
time standpoint when $k$ gets larger, since it materializes the full
marginal and applies randomized response to each cell. 
 
Across all experiments, we conclude that \iht appears to achieve the
best accuracy most consistently, and is very fast in practice. 

\eat{
\parit{$2/3$-way marginals:} $k=2,3$-way marginals each have $4/8$
cells and now-onwards we expect to algorithms performing according to
their design. In marginal based algorithms, we observe that the
performance of \mrs differs a lot from \mps and \mht though the error
bounds proved in theory section for all $3$ algorithms are of the same
order.
For $ k\geq 2$, \mps, \mht start dominating \mrs in most cases. This is because \mrs reports the $1$ at marginal index with much lower probability than \mps ($p_{r}=\frac{3}{4} \times \frac{1}{4} \approx 0.19$ compared to $p_{s}=\frac{3-1}{4+3-1}\approx0.33$ for $e^{\epsilon}=3,k=2$). The difference in the performance of \mps and \mht is tiny and they almost overlap when $k \leq \frac{d}{2}$ confirming the tightness of our analysis. \mps is slightly preferable over \mht for $k=2$ case.
}

\subsection{Impact of privacy parameter $\epsilon$}



\para{Experimental Setting.}
Next, we fix $N$ to $2^{18} \approx 0.25M$ movielens users (sampled
with replacement) and change
$\epsilon$.
As before, we increase $d$ (resp., $k$) along columns
(rows) and vary $0.4 \leq \epsilon \leq 1.4$ to see the effect on
utility in Figure~\ref{fig:vary_eps}.

\para{Observations:} We observe a decline in error as we increase
the privacy budget $\epsilon$.
Once again we see that \ips, \irs, \mrs become unfavorable for $k\geq
2$.
\mps's accuracy gets better than \mht with increase in $\epsilon$,
although \mht is preferable to \mps for small $\epsilon$ values when
$d$ and $k$ are larger.
Yet again, $\iht$ consistently outperforms all other algorithms across all configurations. 
The main takeaway from these experiments is the confirmation that the
algorithms with the best theoretical bounds on performance are borne
out to be the best in practice. 
In general,
\iht is our first preference followed by \mps and \mht.

\subsection{Impact of increasing dimensionality $d$}
\label{sec:Fanti_et_al} 	 

Now that we have established the relative performance of our
algorithms, we compare to an alternative method that works in the  case
$k=2$, denoted \iem (Section~\ref{sec:iem}). 
We consider a larger range of values of the dimensionality $d$,
(achieved by duplicating columns) 
and show the results in Figure~\ref{fig:inp_ht_vs_inp_em_k2}. 
For \iem, we fix the convergence threshold to $\Omega=0.00001$, i.e. stop
when the change in the current guess is below $\Omega$.


\para{Experimental observations.}
We see that the \iem gives reasonable results that improve as
$\epsilon$ is increased.
However, the achieved accuracy is several times worse than the
unbiased estimators \iht and \mps. 
There are additional reasons to not prefer \iem: 
it lacks any accuracy guarantee, and so is hard to predict results.
It is also slow to apply, taking several thousand
or tens of thousands of iterations to converge.
In some cases, the convergance criteria are immediately met by the
uniform distribution, which is far from the true marginal. 
We omit formal timing results for brevity; however, convergence time was observed to grow linearly with $d$.
Weakening the convergence criterion (i.e. increasing the stopping parameter $\Omega$) even
slightly led to much worse accuracy results than the alternative
methods. 
In contrast, our unbiased estimators are found instantaneously. 


\parit{Remark.}
It is reasonable to ask whether EM decoding schemes can be developed for other methods for recovering marginals.
We performed a set of experiments on this approach (details omitted
for space reasons); our conclusion is that while this can be applied
to our algorithms, there is no improvement compared to the
direct construction of unbiased estimators. 

\eat{
 \para{\textbf{EM As An Alternative Decoding Scheme:}} While \iem is not the most preferred solution for $k=2$ case, EM framework in general appears a promising decoding step for reconstructing marginals from noisy data and it is worth exploring whether we can substitute our corrections with tailor made EM based corrections. E.g. we intended to explore whether an EM based scheme can be
 used to decode a joint distribution of noisy Hadamard coefficients? The answer is yes. One can design decoding schemes for all of our algorithms based EM. In the interest of space, we do not detail them. Furthermore, we found performance of such schemes inferior to/at par with their original non-EM based counterparts. Specifically for \iht, replacing our correction with a custom designed EM based correction did not seem to improve \iht's accuracy. 
 }

\section{Applications and Extensions}
Since each cell of a $k$-way marginal is a joint distribution of a set
of $k$ attributes and can be used to determine conditional
probabilities, marginals are useful in machine learning and inference
tasks.
In this section, following our motivational use case, we perform (1)
association testing among attributes (2) dependency trees fitting.
For both  tasks, $1$ and $2$-way marginals are sufficient.
Based on the accuracy results, we use \mps and \iht for these tasks.      
Finally, we discuss how to apply our results to non-binary
attributes. 

\begin{figure}[t]
  \includegraphics[width=\columnwidth]{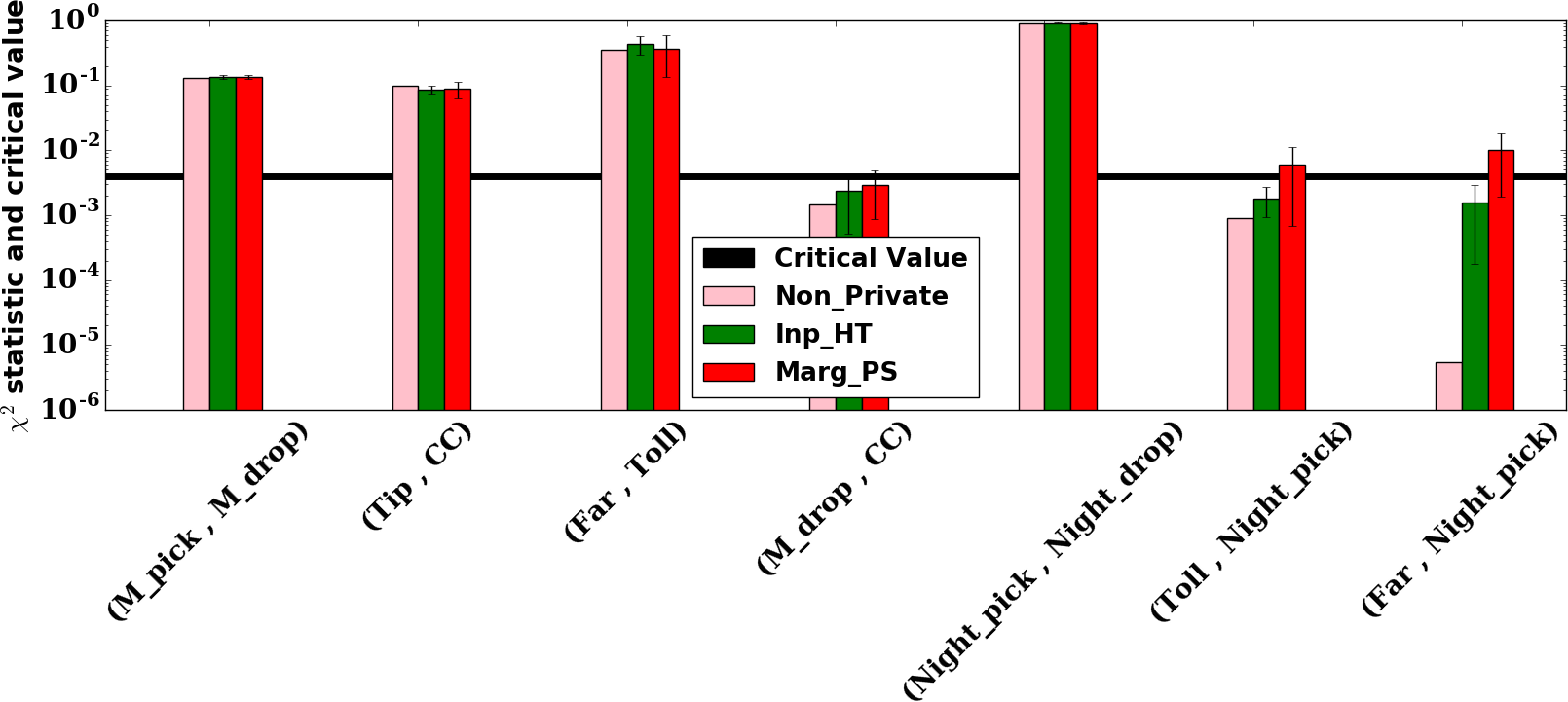}
  \caption{$\chi^{2}$ test values on $N=256K$ NYC taxi trips, $\epsilon=1.1$.}
  \label{fig:chi_2}
\end{figure}

\subsection{Association Testing}
We often want to check if two variables $A,B$ are independent or
not i.e.~we want to know if $\Pr[A,B]\approx \Pr[A]\Pr[B]$.
The $\chi^{2}$ test of independence
compares the observed cell counts to expected counts assuming the
independence (null hypothesis) and compute the $\chi^{2}$ value (see
e.g.~\cite{cda:03}).
It then compares this value to the critical value $p$ for a
given confidence interval (usually $0.95$).
If $\chi^{2} > p$, we conclude that $A,B$ are dependent (rejecting the
null hypothesis).
For a $2$-way marginal $m$, the $\chi^{2}$ statistic is  $\sum_{j \in \{0,1\}^{2}} \frac{(t[j]-\mathbb{E}[t[j]])^2}{\mathbb{E}[t[j]]}$, where $\mathbb{E}[t[j]]$ is the expected value at $t[j]$.

\def\<#1>{$\langle$\ignorespaces#1\unskip$\rangle$}

\para{Experimental setting.}
We use the taxi data for supporting this task since this dataset has a
good mix of correlated/weakly correlated attributes (Figure~\ref{fig:corr}).
As mentioned above, there are strong positive associations in the taxi
data among the pairs \<Night\_pick, Night\_drop >,
\<Toll, Far> and \<CC, Tip> and expect the test to declare them as dependent.
Similarly, we expect the test to declare the pairs \<M\_drop, CC>,
\<Far, Night \_pick > and \<Toll, Night\_pick > to be independent. 

\para{{Experimental observations.}} Figure~\ref{fig:chi_2} compares
privately and non-privately computed $\chi^{2}$ values with the
critical value (computed with $1$ degree of freedom and with
confidence interval of $95\%$\footnote{Gaboardi et
al.~in \cite{grv:16} suggest increasing $p$ since
comparing a differentially private $\chi^{2}$ statistic to a noise
unaware critical value may not lead to a good significance level even
for large $N$.  We do not perform correction in this test, and leave developing
robust correlation tests under LDP for future work.})
over log scale. We observe that non-private and private  $\chi^{2}$
values are quite close in most cases for \iht (note the log scale on
the y-axis, which tends to exaggerate errors in small quantities).
On the other hand, \mps often commits the type $I$ error (thus failing
to reject the null hypothesis) for the pairs \<Toll, Night\_pick >, \<
Far, Night\_pick > and occasionally for pairs \<M\_drop , CC>, since
the test statistic is close to the critical value in these cases. 

\begin{figure}[t]
  \includegraphics[width=\columnwidth]{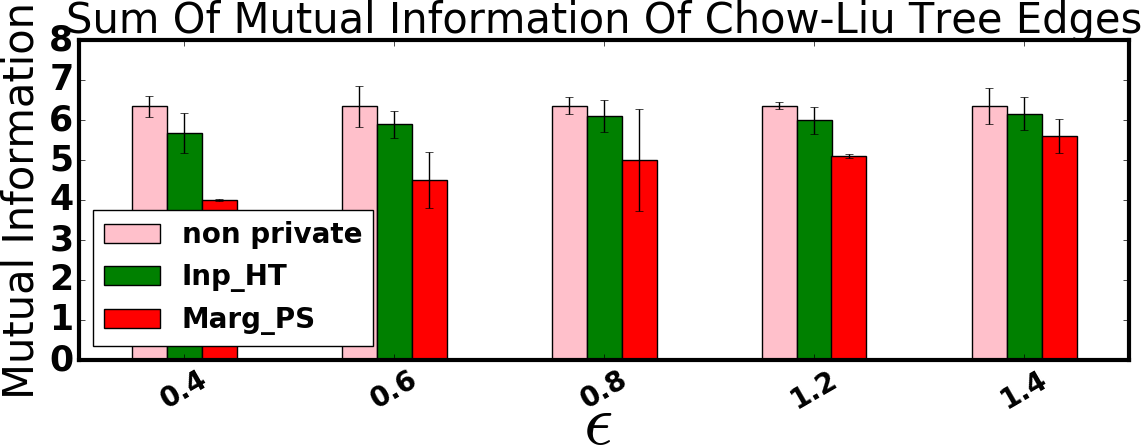}
  \caption{Total mutual information of trees on movielens}
  \label{fig:chow_liu_tree}
\end{figure}

\subsection{Bayesian Modeling} \label{sec:chow_liu}
Exact estimation of a joint distribution for $d$ discrete variables
could be computationally infeasible for large $d$'s.
Chow and Liu in \cite{cl:68} proposed an algorithm for approximating a
joint distribution of a set of discrete variables using products of
distributions involving no more than pairs of variables.
Since each variable in the approximation depends on at most one more
variable, the task of finding such approximation can be thought as
finding a tree that optimizes a particular distance metric.
They prove that a tree configuration that maximizes total mutual
information among edges is an optimal approximation of the joint
distribution in question. This insight converts the intractable
optimization problem of finding such tree to an easy problem of
finding a maximum weight spanning tree.
Concretely, all we have to do is treat all random variables as nodes
in an empty graph and find a tree that maximizes the total edge
weight.
Once a tree is learnt, any high dimensional joint distribution of interest can be learnt by multiplying conditional probabilities that can found using marginals. 

The center piece of this algorithm is computation of mutual information between ${d\choose 2}$ pairs of variables. Mutual information between two discrete variables $A,B \in \{0,1\}$ is given as
\[ \textstyle
 MI(A,B)=\sum_{i,j \in \{0,1\}^2}\Pr[A=i,B=j]\log \frac{\Pr[A=i,B=j]}{\Pr[A=i]\Pr[B=j]}
 \]	

\para{Experimental setting.}
Note that the Chow-Liu algorithm finds a tree from the equivalence
class of trees fitting the given data and are not unique.
Moreover, there could be many others trees with different topologies
achieving near optimal MI score.
Therefore, our aim in this section is to compare  total MI from
privately and non-privately learnt trees.
For this purpose, we use the movielens dataset with $d=10$.

\para{{Experimental observations.}}
Figure~\ref{fig:chow_liu_tree} compares the total (true) MI from $200K$ users
for various $\epsilon$ values (error bars show variation over
different subsets of sampled records). 
We once again see that MI of trees computed with \iht marginals is
nearly the same as the non-private computation.
\mps is less accurate at low $\epsilon$'s but catches up with \iht as $\epsilon$ increases.
We conclude that \iht gives a robust solution for this approach. 

%
%
%

\subsection{Categoric Attributes}

We now consider how to apply these methods over more general
classes of input -- in particular, over cases where the input is
non-binary, but ranges over a larger set of possible categories $r>2$.
Suppose now we have $d$ categoric attributes with cardinalities
(indexed in order of size for convenience)
$r_1 \ge  r_2 \ge \ldots \ge r_d$, and wish to find marginals
involving subsets of at most $k$ attributes. 
We describe two approaches to handling such data. 

\para{Binary encoding methods using our algorithms.}
Many of our algorithms such \mrs, \mps, \ips, \irs will generalize
easily in this case, since they can be applied to users represented as
sparse binary vectors.
The Hadamard-based methods \mht and \iht can also be generalized if we
rewrite the input in a binary format, i.e. we create a fresh binary
attribute for each possible categoric value in an attribute (aka
``one-hot encoding'').
However, we can more compactly encode an attribute value that takes on
$r$ possible values using $\lceil \log_2 r \rceil$ bits, and consider
this as the conjunction of $\lceil \log_2 r \rceil$ binary
attributes.
Consequently, we state a result (based on our strongest algorithm for
the binary case) in terms of the effective binary
dimension of the encoded data,
$d_2 = \sum_{i =1}^{d} \lceil \log_2 r_i \rceil$; and the binary
dimension of $k$-way marginals
$k_2 = \sum_{i=1}^{k} \lceil \log_2 r_i \rceil$:

\begin{corollary}
  Using \iht on binary encoded data, we achieve $\epsilon$-LDP, and
  with constant probability we have for any target $k$-way marginal
  $\beta$ on
  binary encoded data,\newline
\centerline{$ \|C^\beta(t) - C^{\beta}(t^*)\|_1 = \tilde{O}\left( \frac{(2d_2)^{k_2/2}}{\epsilon\sqrt{N}}\right)$}
\end{corollary}

Consequently, this provides an effective solution, particularly for
data with low cardinality attributes.  

\para{Orthogonal Decomposition.}
It is natural to ask whether there are alternative decompositions
for categorical data which share many of the properties of the
Hadamard transform (orthogonal, requiring few coefficients to
reconstruct low-order marginals). 
One such approach is the \textit{Efron-Stein decomposition}
\cite{es:81} which is a generalization of Hadamard transform for non
binary contingency tables. Similar to HT, it is possible to extract a
set of Efron-Stein coefficients necessary and sufficient to evaluate a
full set of a $k-$way marginals.
One could then design an algorithm similar to \iht that adds noise to
a random coefficient, allowing an unbiased estimate to be constructed
by an aggregator.
We conjecture that for low order marginals, a scheme based on such decomposition will be among the best solutions. 

\section{Concluding Remarks}
\label{sec:nonbinary}

We have provided algorithms and results for the central problem of
private release of marginal statistics on populations.
Our main conclusion is that methods based on Fourier (Hadamard)
transformations of the input are effective for this task, and have
strong theoretical guarantees.
Although the technical analysis is somewhat involved, the algorithms
are quite simple to implement and so would be suitable for inclusion in
current LDP deployments in browsers and mobile devices: it would
require only small modifications to RAPPOR or iOS to incorporate them.

\clearpage
\begin{appendix}

\begin{algorithm}[t]
\caption{\iht -- User Routine}
\begin{algorithmic}[1]
\Procedure{\iht}{$t_i$}
\State Let $j_i$ be the signal index of $t_i$.
\State Randomly sample a coefficient index  $\ell_i \in H_k$.
\State $h_i \gets (-1^{\langle j_i \wedge \ell_i \rangle})$
\Comment{Compute (scaled up) $\theta_{\ell_i}$}
\State $t^{*}_i \gets RR(h_i[j_i])$
\Comment{Apply randomized response}
\State Return ($t^{*}_i,j_i$)

\EndProcedure
\end{algorithmic}
\label{alg:iht_user}
\end{algorithm}

\begin{algorithm}
\caption{Aggregator's routine for \iht}
\begin{algorithmic}[1]


\State 	$T^{*}[0]= 1$ \Comment{0th Hadamard coefficient is always $1$.}
\State \mbox{Aggregator populates $(t^{*}_i,\ell_i)$ from all users in
array $T^{*} \in \mathbb{R}^{|H_{k}|}$.}
\ForAll{$j \in H_{k}$}
\State $T^{*}[j] \gets\frac{\sum_{i=1}^N T_i^{*}[j]}{N_j (2p-1)}$. \Comment{$N_j$ is the frequency count of index $j$.}
\EndFor
\State Use Lemma~\ref{fact:ht} on $T^{*}$ to evaluate any $\leq k-$way marginal.

\end{algorithmic}
\label{alg:input_aggr}
\end{algorithm}

\eat{
\begin{figure*}
\caption{Landscape Of Methods}
\centering
\includegraphics[width=\textheight,angle=90]{method_basis.pdf}
\label{fig:landscape}
\label{fig:method_remarks} 
\end{figure*}  
}

\section{Deferred Proofs}
  
In this appendix, we provide the detailed technical proofs for the
claimed privacy and accuracy properties of our algorithms.
Note that while the proofs are somewhat involved, the algorithms
themselves are fairly straightforward to implement.
By way of example, in Algorithms~\ref{alg:iht_user}
and \ref{alg:input_aggr} we show the procedure followed for user and
aggregator for the \iht approach, which is perhaps the most involved.
The user's work is straightforward, as they just have to sample a
coefficient from the set of relevant coefficients $H_k$, and apply
randomized response.
The aggregator tallies these responses, applies bias correction and
normalizes, then treats these as if they were the exact coefficients
to evaluate any marginal with (following the equation stated in
Lemma~\ref{fact:ht}). 

\begin{proof}[Proof of Theorem~\ref{thm:master}]
 We first consider the input of a single user subject to
 randomized response, and obtain an unbiased estimate for their
 contribution to the population statistics.
 This lets us combine the estimates from each user to compute an unbiased estimate for the population, whose variance we analyze
to bound the overall error. 
  
Let $t_i[j] \in \{-1,1\}$ be $i$'s unknown true input at location $j$
and $t^{*}_i[j]$ be the unbiased estimate of $t_i[j]$. 
First, we derive the values we should ascribe to $t^*$ to ensure
unbiasedness, i.e.~$\E[t^{*}_i[j]]=t_i[j]$.

\medskip
\noindent
1. When $j$ is sampled (with probability $p_s$) and $t_i[j] = 1$, 
  we set $t^{*}_i[j] = x/p_s$ with probability $p_r$
  and $t^*_i[j] = y/p_s$ otherwise. 

\smallskip  
\noindent
2. When $j$ is sampled (with probability $p_s$), and $t_i[j] = -1$, 
  we set $t^{*}_i[j] = y/p_s$ with probability $p_r$ and $x/p_s$ otherwise.

\smallskip  
\noindent
3. When $j$ is not sampled, we implicitly set $t^{*}_i[j] = 0$. 

\smallskip
We can encode these conditions with linear equations: 
\begin{align}
p_rx +(1-p_r)y=-1
\\
p_ry+(1-p_r)x=1
\end{align} 
Solving, we obtain $x=\frac{1}{(2p_r-1)}$ and $y=-\frac{1}{(2p_r-1)}.$
As we require $p_r > \frac12$, we have $x >0$ and $y = -x <0$.
We now analyze the (squared) error from using these parameters. 
Define a random variable for the observed error as $Y_i[j]=t^{*}_i[j]-t_i[j]$.
Observe that $\E[Y_i[j]]$ is 0, and
\[|Y_i[j]| \leq \frac{1}{p_s}\left(1 + \frac{1}{2p_r - 1}\right) =
\frac{2p_r}{p_s(2p_r - 1)} := M.\]
Furthermore, $|Y_i[j]|$ is symmetric whether $t_i[j] = 1$ or $-1$.
Then:
\begin{align}
\nonumber
  \Var&[Y_i[j]]  = \E[Y_i^{2}[j]] \\
\nonumber
  &=  \frac{p_rp_s}{p_s^2}\big|{ \frac{1}{2p_r-1}-1}\big|^{2}
  +\frac{(1-p_r)p_s}{p_s^2}\big|{1+\frac{1}{2p_r-1}}\big|^{2}
  + (1-p_s)1^2\\
\nonumber
& \le \frac{p_r}{p_s}\left(\frac{2p_r - 2}{2p_r-1}\right)^2 + \frac{(1-p_r)}{p_s}\left(
  \frac{2p_r}{2p_r-1}\right)^2 + (1-p_s)\\
\nonumber
  & = \frac{4}{p_s(2p_r-1)^2} ( p_r (1-p_r)^2 + (1-p_r)p_r^2) +(1-p_s)\\
& = \frac{4p_r(1-p_r)}{p_s(2p_r-1)^2} +(1-p_s) := \sigma^2.
\label{eq:varbound}
\eat{
  =p(x-1)^{2}+(1-p)(1+x)^{2}\\
  =p(x^{2}-2x+1)+(1-p)(1+2x+x^{2})\\
  =1+2x-4xp+x^{2}\\
  =(1+x)^{2}-4px\\
  =(1+\frac{1}{2p-1})^{2}-\frac{4p}{2p-1}\\
  =(\frac{2p}{2p-1})^{2}-\frac{4p}{2p-1}\\
  =(\frac{4p}{2p-1})(\frac{1-p}{2p-1}) 
}
\end{align}


Now we consider the effect of aggregating $N$ estimates of the
$j$'th population parameter.
Using Bernstein's inequality (Definition~\ref{def:bernstein}), we can bound the probability of the
error being large based on the bound $M$ on the absolute value of the
$Y_i[j]$'s. 


\begin{align}
\nonumber
  \Pr& \left[\frac{|\sum_{i=1}^N Y_i[j]|}{N} \geq c\right] \leq 2\exp
  \left(-\frac{Nc^{2}}{2\sigma^2+\frac{2cM}{3}}\right)
  \\
  \nonumber
& \le 2\exp
  \left(-\frac{Nc^{2}}{2(\frac{p_r(1-p_r)}{p_s(2p_r-1)^{2}} +1)+\frac{2c
      p_r}{3p_s(2p_r-1)}}\right)\\
  &
  =2\exp  \left(-\frac{Nc^{2}}{\frac{2p_r}{p_s(2p_r-1)}(\frac{2(1-p_r)}{(2p_r-1)}+\frac{c}{3})
  +2}\right)
\label{eq:master}
\end{align}

This provides us with the statement of the theorem. 
\end{proof}

\begin{proof}[Proof of Theorem~\ref{thm:irs}] 
We first analyze the accuracy with which each entry of the full
marginal $t[j]$ is reconstructed, then combine these to obtain the
overall result. 
Consider an arbitrary index $j \in 2^{d}$, since \irs is symmetric across all indices.
To achieve $\epsilon$-LDP, we set
$p_r=\frac{e^{\epsilon/2}}{1+e^{\epsilon/2}}$, and $p_s=1$.
For the purpose of analysis only, 
we reduce the problem so that we can apply
Theorem~\ref{thm:master}, by applying a remapping from $\{0,1\}$ to $\{-1,1\}$:
we replace $t_i[j]$ with $t'_i[j] = 2t_i[j] - 1$. 
Observe that the absolute error in reconstructing $t'_i[j]$ is only a
constant factor of that in reconstructing $t'_i[j]$. 
%
Writing $\alpha = e^{\epsilon/2}$, 
then we have the variance of the local errors $Y_i[j] = (t_i[j] - t^*_i[j])$ is (substituting
these values of $p_r$ and $p_s$ into~\eqref{eq:varbound}):
\begin{align*}
  \Var[Y_i[j]] & \le 4\frac{p_r(1-p_r)}{(2p_r-1)^2}+1-1
  =4\frac{(\frac{1}{1+\alpha})(1-\frac{1}{1+\alpha})}{(\frac{2}{1+\alpha}-1)^2}\\
  & = 4\frac{\frac{\alpha}{(1+\alpha)^2}}{(\frac{1-\alpha}{1+\alpha})^2}
  =\frac{4\alpha}{(1-\alpha)^2}
  =\frac{4e^{\epsilon/2}}{(e^{\epsilon/2}-1)^2}.
\end{align*}

The reconstruction of the full input distribution is
$t^* = \sum_{i=1}^N t^*_i/N$. 
We can make use of the inequalities
$\frac{1}{e^{\epsilon/2}-1} \le \frac{1}{\epsilon}$ and
$1 <  e^{\epsilon/2} < 4$ for $0 < \epsilon < 2$ to bound the variance
and substitute into~\eqref{eq:master}.
\[
\Pr[|t_j - t^*_j| > c]
\le 2 \exp\bigg(- \frac{Nc^2}{2\cdot(4 \frac{8}{\epsilon^2}) +
  \frac{2\cdot8c}{3\epsilon}}\bigg)
\]

\noindent
Setting $c$ to $9 N^{-1/2}\frac{1}{\epsilon}
\sqrt{\log 2^{d+1}/\delta}$
bounds this probability to
\begin{align*}
& 2\exp\Bigg( -\frac{81 \frac{1}{\epsilon^2} \log 2^{d+1}/\delta}{
  \frac{32}{\epsilon^2}  + \frac{16}{3} \frac{9}{\epsilon^2}
  \sqrt{\frac{2^d \log 2^{d+1}/\delta}{N}}}\Bigg)
\\
<&
  2 \exp\Big(-\frac{81\log(2^{d+1}/\delta)}{32 + 48}\Big)
  \leq \delta/2^d
\end{align*}

\noindent
This ensures that this error probability is less than
$\delta/2^d$ for any index $j$.
This limits the error in each of the $2^d$ estimates to being
$\tilde{O}(\frac{1}{\epsilon} \sqrt{\frac{1}{N}})$, by
applying a union bound.

We construct the target marginal $\beta$ via the marginal operator, so
$\widehat{\mathcal{C}^\beta} = \mathcal{C}^{\beta}(t^*)$.
Each entry $t^*[j]$ is an unbiased estimator for $t[j]$ whose absolute
value is bounded by $c$ with probability $1-\delta$.
Conditioning on this event,
we compute $\widehat{\mathcal{C}^\beta}[\gamma] =
\sum_{\alpha \preceq \gamma} t^*[\alpha]$, summing over the
$2^{d-k}$ values of $\alpha \preceq \gamma$.
The error in this quantity is then at most
$\tilde{O}(c\sqrt{2^{d-k}})$, applying a Hoeffding bound (Definition~\ref{def:hoeffding}). 
Finally, summing the absolute errors over all $2^k$ entries $\gamma$ in the
target marginal $\beta$, we have probability at least
$1-\delta$ that the total variation distance is
$\tilde{O}(\frac{2^{k}2^{(d-k)/2}}{\epsilon\sqrt{N}})
= \tilde{O}(\frac{2^{(d+k)/2}}{\epsilon\sqrt{N}})$. 
%
%
\eat{
Using Bernstein's inequality: 
\begin{align*}
E_{max}=\underset{j \in [2^{d}]}{\operatorname{max}} Pr[\frac{\sum_{i=1}^N|t_i^{*}[j] -t_i[j]|}{N} \geq c] & \leq 2\exp \Big(-\frac{Nc^{2}}{2(O(\frac{1+\epsilon}{\epsilon^{2}})+\frac{c}{3})}\Big)  
 \end{align*}
Using $c =\mathcal{O}\Big(\sqrt{\frac{2^{d}(1+\epsilon)\ln(\frac{2^{d}}{\tau})}{\epsilon^{2}N}}\Big)$, we get a W.H.P bound. 
$E_{total} \leq2^{d} E_{max}$.
For $\epsilon \in [0,1], \mathcal{O}(\frac{1+\epsilon}{\epsilon^{2}})
\leq \mathcal{O}(\frac{1}{\epsilon^{2}}), E_{max} \leq
\mathcal{O}(\frac{2^{3d/2}}{\epsilon}\sqrt{\frac{\ln(\frac{2^{d}}{\tau})}{N}})$. For
$\epsilon >1, E_{max} \leq
\mathcal{O}(2^{3d/2}\sqrt{\frac{\ln(\frac{2^{d}}{\tau})}{\epsilon
    N}})$
}
\end{proof}

\begin{proof}[Proof of Theorem~\ref{thm:psmaster}]
Similar to Theorem~\ref{thm:master}, we define random variables
$Y_i[j]$ which describe the error in the estimate from user $i$ at
position $j$.
The proof is a bit more complicated here, since these variables are not
symmetric. 
Consider user $i$ who samples a location under PS, such that the correct location is
sampled with probability $p_s$, and each of the $D=2^d-1$ incorrect
locations is sampled with probability $(1-p_s)/D$.
Following the analysis in Section~\ref{sec:accuracy},
we report $\frac{D + p_s - 1}{Dp_s + p_s - 1}$ for the location which
is sampled, and $\frac{p_s - 1}{Dp_s + p_s - 1}$ for those which are
not sampled.
For convenience, define the quantity $\Delta = {Dp_s + p_s - 1}$.
The choice of $p_s$ (which depends on $D$ and $\epsilon$) ensures that
$\Delta > 0$. 
There are two cases that arise:

\smallskip
\noindent
(i) $t_i[j] = 1$.
With probability $p_s$, location $j$ is sampled.
The contribution to the error at this location is \newline
$\frac{D + p_s - 1}{\Delta} - 1
= \frac{1}{\Delta}(D + p_s - 1 - Dp_s - p_s +1)
= \frac{D}{\Delta}(1-p_s)
$.

\smallskip
\noindent
Else, with probability $1-p_s$, $j$ is not sampled, generating
error\newline
$ \frac{p_s -1}{\Delta}-1
= \frac{p_s-1-Dp_s-p_s+1}{\Delta}
= \frac{D}{\Delta}p_s
$
for $|t_i^*[j] = t_i[j]|$. 

\noindent
\smallskip
(ii) $t_i[j]=0$. 
With probability $\frac{1-p_s}{D}$, we sample this $j$, 
giving error $\frac{D + p_s -1}{\Delta} - 0$.
Otherwise, the contribution to the error is
$\frac{p_s - 1}{\Delta}$.


\medskip
We define a random variable $Y_i[j]$, which is the error resulting from
user $i$ in their estimate of $t_i[j]$.
Note that an upper bound $M$ on $Y_i[j]$ is $D/\Delta$. 
We compute bounds on $Y_i^2$, conditioned on $t_i[j]$.
\begin{align*}
  \E[Y_i[j]^2 | t_i[j] = & 1]  =
p_s\Big(\frac{D}{\Delta}(1-p_s)\Big)^2 + (1-p_s)\Big(p_s\frac{D}{\Delta}\Big)^2
\\
& = p_s(1-p_s)\Big(\frac{D}{\Delta}\Big)^2
\leq (1-p_s) \frac{D^2}{\Delta^2}\\
%
  \E[Y_i[j]^2 | t_i[j]=& 0] 
  = \frac{1-p_s}{D}\left(\frac{D + p_s - 1}{\Delta}\right)^2 +
 \left(1-\frac{1-p_s}{D}\right)\left(\frac{p_s-1}{\Delta}\right)^2
 \\
 &
 = \frac{1-p_s}{\Delta^2}\left(\frac{1}{D}(D + p_s - 1)^2 + \frac{D + p_s -
   1}{D}(1-p_s)\right)
 \\
 &
 = \frac{1-p_s}{D\Delta^2}(D + p_s -1)(D + p_s - 1 + 1 - p_s)
 \\
 & = (1-p_s)(D + p_s -1)/\Delta^2 \leq (1-p_s)D/\Delta^2
  \end{align*}

To bound the error in $t^*[j]$, we make use of the (unknown) parameter $f_j$, the
proportion of users for whom  $t_i[j]=1$.
We subsequently remove the dependence on this quantity.
We now write 
\[
\E[Y_i[j]^2] \leq (1-p_s)\frac{D}{\Delta^2}(f_j D + (1-f_j)) := \sigma_j^2\]

Using this in the Bernstein inequality (Definition~\ref{def:bernstein}), we obtain
\begin{align*}
\Pr&
\left[\frac{|\sum_{i=1}^{N} Y_i[j]|}{N}\!\geq\!c_j\right]  \leq
2\exp\left(-Nc_j^2\big/\left(2\sigma_j^2+\frac{2c_jM}{3}\right)\right)
  \\
  &
= 2\exp\left( -\frac{Nc_j^2}{2(1-p_s) \frac{D}{\Delta^2}(f_j D +
  (1-f_j)) + \frac{2c_jD}{3\Delta}}\right)
  \end{align*}
If we write $\Psi_j = \sqrt{f_j D + 1-f_j}$, 
then setting
$ c_j = \frac{\sqrt{3D\ln (2/\delta)}}{\Delta\sqrt{N}} \Psi_j$
is sufficient to ensure that this probability is at most $\delta$.%
\eat{
To verify this, observe that substituting this choice of $c_j$ yields
\[
\Pr[{
\frac{|\sum_{i=1}^{N} Y_i[j]|}{N}}\geq c_j] 
\leq 2 \exp\left( - \frac{\ln (2/\delta)}{(1-p_s) + 2c_j\Delta/3\Psi_j^2}\right)\]

With our assumption that  $N \geq D \ln (1/\delta)$,
and observing $\Psi_j > 1$, 
then this probability is at most
\[ 2\exp\left( - \frac{\ln (2/\delta)}{\frac{2}{3}(1-p_s) + 2/(9\Psi_j)}\right)
\leq 2\exp( -\frac{\ln(2/\delta)}{2/3 + 2/9})
< \delta
\]
}%
When we apply the marginal operator $\mathcal{C}^\beta$ to the
reconstructed input $t^*$, each of the $2^k$ entries is formed by
summing up $(D+1)/2^k$ (unbiased) entries of $t^*$.
Write $f'_\gamma = \sum_{j \wedge \beta = \gamma} f_j$, 
and define $\Psi'_\gamma$ correspondingly
as $\sqrt{f'_\gamma(D-1) + \frac{D+1}{2^k}}$. 
Applying the Hoeffding bound (Definition~\ref{def:hoeffding}),
we obtain that each
$\mathcal{C}^\beta(t^*)[\gamma]$ has error at most
$\frac{\sqrt{3D\ln(2/\delta)}}{\Delta\sqrt{N}} \Psi'_\gamma$
with probability at least $1-\delta$.  

We can now sum the error across all $(D+1)/2^k$ indices $\gamma$.
First, 
\begin{align*}
\sum_{\gamma \preceq \beta} \psi'_\gamma
& =
\sum_{\gamma \preceq \beta} (f'_\gamma (D-1) + \frac{D+1}{2^k})^{\frac12}
\\\leq&
  \sqrt{2^k} \Big(\sum_{\gamma\preceq\beta} f'_\gamma (D-1) + \frac{D+1}{2^k}\Big)^{\frac12}
  =\sqrt{2^{k+1} \cdot D} 
  \end{align*}

\noindent
where the inequality is due to Cauchy-Schwarz,
  and we use that the $f'_\gamma$s are a probability distribution, and sum
  to 1. 
Then
we have a bound on the total variational error error of marginal construction by summing over
all indices $\gamma$ as
\[
  \sum_{\gamma\preceq\beta} \frac{c'_\gamma}{2}  =
  \frac{1}{2\Delta} \sqrt{\frac{D}{N}} \sqrt{3\ln 2/\delta}
  \sum_{\gamma\preceq\beta} \Psi'_\gamma 
   \leq 
  \frac{2^{k/2}D}{\Delta\sqrt{N}} \sqrt{\textstyle\frac32 \ln 2/\delta}
  \]
\eat{
Computing the expected squared error over the whole marginal,
we obtain

\begin{align*}
  & \frac{1}{\Delta^2}\Big(p_sD^2(1-p_s)^2 + (1-p_s)D^2p_s^2 \\
  & + D\frac{1-p_s}{D}(D + p_s - 1)^2
  + D(1 - \frac{1-p_s}{D})(p_s - 1)^2\Big)
  \\
  = & \frac{1}{\Delta^2}(D^2p_s(1-p_s)
  + (1 - p_s)(D + p_s-1)^2 + (D - 1 + p_s)(p_s-1)^2)
  \\
  = & \frac{1}{\Delta^2}(D^2p_s(1-p_s)
  + (1-p_s)(D + p_s -1)(D + p_s - 1 + 1 - p_s)) \\
  = & \frac{D(1-p_s)}{\Delta^2}(Dp_s + D + p_s - 1)\\
  = & \frac{D(1-p_s)}{\Delta^2}(\Delta + D) \\
  = & \frac{D}{\Delta}(1 + \frac{D}{\Delta})(1-p_s)
\end{align*}
}
We next  simplify the term 
$D/\Delta$ as follows.
Recall that theory sets $p_s = (1 + De^{-\epsilon})^{-1}$.
Then

\begin{align*}
\frac{D}{\Delta} & = \frac{D}{(D+1)/(1 + De^{-\epsilon}) - 1} 
  = \frac{D(1+De^{-\epsilon})}{D+1 - 1 - De^{-\epsilon}} \\
&  = \frac{1 + De^{-\epsilon}}{1 - e^{-\epsilon}}
= \frac{1}{1-e^{-\epsilon}} + \frac{D}{e^{\epsilon}-1}
\end{align*}

When $D$ is very small, in particular when $D=1$, this reduces to a
similar error as for the RR
case.
Assuming that $\epsilon$ is at most a constant (say, 8),
we can upper bound this expression by
$O(\frac{D+1}{\epsilon})$. 
%
%
Hence, the total variational error is bounded by
$\tilde{O}(\frac{2^{k/2}(D+1)}{\epsilon\sqrt{N}})$. 
\end{proof}

\begin{proof}[Proof of Theorem~\ref{thm:iht}]
The proof proceeds along the same lines as for Theorem~\ref{thm:irs}.
We set $p_r = e^{\epsilon}/(1 + e^{\epsilon})$ to ensure
  that \iht meets $\epsilon$-LDP.
Recall that, from Lemma~\ref{fact:sumcoeffs}, our aim is to compute Hadamard
coefficients as the normalized sum of the coefficients from each
user. 
To apply the Master theorem (Theorem~\ref{thm:master}), we first multiply up
each coefficient by the
$2^{d/2}$ factor from the Hadamard coefficients $\theta$
(Definition~\ref{def:ht}).
Since each user's input vector has only a single 1 entry, this ensures
that each $\theta_i[j]$ is either $-1$ or $+1$. 
Now the $\theta_i$ and $\theta_i^*$s represent the $T$ necessary and
sufficient (scaled up) Hadamard coefficients, and so we set $p_s= 1/T$. 
We write the variance of the errors in these estimates $Y_i[j]$,
and obtain
\[
    \Var[Y_i[j]] = 4T\frac{p_r(1-p_r)}{(2p_r-1)^2} + 1 =
    \frac{4Te^\epsilon}{(e^\epsilon - 1)^2} + 1 = O(T/\epsilon^2)
 \]

 Substituting this variance bound into \eqref{eq:master}, we obtain \newline
%
\[
 \Pr\Big[ \frac{|\sum_{i=1}^{N} Y_i[j]|}{N} \ge c \Big] \leq 2
 \exp\left(-\frac{Nc^2}{O(T/\epsilon^2 + \frac{Tc}{\epsilon})}\right)
\]
 Setting $c$ proportional to $N^{-1/2} \frac{1}{\epsilon}{\sqrt{T\cdot
     \log {T/\delta}}}$
 ensures that this probability is at most $\delta/T$ for any given
 Hadamard coefficient $j$ (again using that $N$ is large enough).
This bound then holds for all $T$ with
 probability $1-\delta$, using the union bound. 

 In order to translate this into a bound on the accuracy of
 reconstructing a marginal, we make use of Lemma~\ref{fact:ht}, that
 the marginal can be expressed in terms of a linear sum of Hadamard
 coefficients.
 Adapting~\eqref{eq:htsum}, we have that
\begin{align*}
\sum_{\gamma \preceq \beta}|\mathcal{C}^\beta[\gamma] -
\widehat{\mathcal{C}^\beta}[\gamma] |
\leq
\sum_{\gamma \preceq \beta} \big|
\sum_{\alpha \preceq \beta} (\theta_\alpha - \hat{\theta}_\alpha)
\sum_{\eta \wedge \beta = \gamma} \phi_{\alpha,\eta}~\big|
\end{align*}
\noindent
To bound this quantity, we observe that: \newline
(i) There are $2^k$ such $\gamma \preceq \beta$ to consider.
\newline
(ii) There are similarly $2^{k}$ such $\alpha$ to consider,
and the above analysis bounds
$(\theta_\alpha - \hat{\theta}_\alpha) \leq c/2^{d/2}$,
once we rescale the coefficients back down.
Since the $\hat{\theta}_\alpha$ are unbiased estimators bounded by $c2^{-d/2}$, by the
Hoeffding inequality, we have that the sum of $2^{k}$ of these is 
at most $2^{k/2-d/2}c$ with probability at least $1-\delta$.
\newline
(iii)
Given $\gamma \preceq \beta$, there are $2^{d-k}$
such $\eta$ to consider, and so we have
$|\sum_{\eta \wedge \beta = \gamma} \phi_{\alpha,\eta}| \leq
2^{d-k}2^{-d/2} = 2^{d/2 - k}$.

Then the total variational error is (multiplying these three
quantities together)
$2^{k} 2^{k/2-d/2} c 2^{d/2 - k} = c2^{k/2} =
\tilde{O}\left(\frac{2^{k/2} \sqrt{T}}{\epsilon\sqrt{N}}\right)$. 
\eat{
The expression $\sum_{\eta: \eta \cap \alpha = \gamma}
 \phi_{\alpha,\gamma}$ in \eqref{eq:htsum}
 is only non-zero when $\alpha = \gamma$
 (\cite[Theorem 5]{barak:07}).
 There are $2^{d-k}$ values of $\eta$ which satisfy the condition, and
 each contributes a weight of $2^{-d/2}$, so the weight placed on
 $\theta_\alpha$ is $2^{d/2 - k}$.
 Summed over the $2^k$ values of $\alpha$ that satisfy
 $\alpha \preceq \beta$, we obtain a (linear) weight of $2^{d/2}$
 placed on the Hadamard coefficients.
 However, this factor of $2^{d/2}$ cancels when we properly rescale
 the Hadamard coefficients according to Definition~\ref{def:ht}. 
 Thus, the total (scaled) error in the reconstruction of the marginal
 is at most
 $N^{-1/2} \frac{1}{\epsilon}\sqrt{T \cdot \log {T/\delta}}$.
}
%
\end{proof}

\begin{figure}[t]
\centering
\begin{tabular}{|c|c|c|c|c|}
\hline 
N & d & k & $\epsilon$ & Failed/Total Marginals \\
\hline 
$2^{16}$ & 8 & 1 & 0.2 & 3/8 \\  
\hline
$2^{18}$ & 8 & 2 & 0.1 & 15/28 \\  
\hline
$2^{16}$ & 8 & 2 & 0.2 & 3/28 \\
\hline 
 
$2^{16}$ & 12 & 2 & 0.2 & 19/66 \\

\hline 
$2^{18}$ & 16 & 2 & 0.1 & 120/120 \\ 
\hline 
$2^{18}$ & 16 & 2 & 0.2 & 72/120 \\ 
\hline 
$2^{19}$ & 24 & 2 & 0.2 & 276/276 \\ 
\hline 
\end{tabular}
\caption{Failure rate for \iem on NYC taxi data for small $\epsilon$
values}. 
\label{fig:EM_fail}
\end{figure}

\end{appendix}

\clearpage
\bibliographystyle{abbrv}
\bibliography{papers.bib}

\end{document}